\documentclass[12pt,preprint]{aastex}

\usepackage{epsfig}

\shorttitle{Line survey of CIT\,6}
\shortauthors{Zhang, Kwok, \& Nakashima}

\begin{document}

\title{A molecular line survey of the extreme carbon star
CRL\,3068 at millimeter wavelengths}

 \author{Yong Zhang, Sun Kwok, and Jun-ichi Nakashima}
 \affil{Department of Physics, University of Hong Kong, Pokfulam, Hong Kong}
 \email{zhangy96@hku.hk; sunkwok@hku.hk; junichi@hku.hk}

\begin{abstract}

We present the results of a molecular line survey of the extreme carbon star \object{CRL\,3068}.  The observations were carried out with the Arizona Radio Observatory (ARO) 12\,m telescope and the Heinrich Hertz Submillimeter Telescope (SMT)  at the $\lambda$ 2\,mm and $\lambda$ 1.3\,mm atmospheric windows. The observations
cover the frequency bands from 130--162\,GHz and 219.5--267.5\,GHz.
The typical sensitivities achieved are $T_R<15$\,mK and $T_R<7$\,mK for the ARO 12\,m and SMT, respectively.  Seventy two individual emission features belonging to 23 molecular species and isotopologues were detected.  Only three faint lines remain unidentified. The species $c$-C$_3$H, CH$_3$CN, SiC$_2$, and the isotopologues, C$^{17}$O and C$^{18}$O, HC$^{15}$N, HN$^{13}$C, C$^{33}$S, C$^{34}$S, $^{13}$CS, $^{29}$SiS, and $^{30}$SiS are detected in this object for the first time.
Rotational diagram analysis is carried out to determine the column densities and excitation temperatures. The isotopic ratios of the elements C, N, O, S, and Si have also been estimated. 
The results are consistent with stellar CNO processing and suggest that \object{CRL\,3068}  is more carbon rich than \object{IRC+10216} and \object{CIT\,6}. 
It is also shown that the chemical composition in \object{CRL\,3068} is somewhat different from that in \object{IRC+10216} with a more extensive synthesis of  cyclic and long-chain molecules in \object{CRL\,3068}.
The results will provide valuable clues for better understanding circumstellar 
chemistry.

\end{abstract}


\section{Introduction}

As a star evolves up the asymptotic giant branch (AGB), copious amount of material at its outer layers
are ejected into space through stellar winds and form an expanding circumstellar envelope. 
Through millimeter-wave and infrared spectroscopy, we have learned that circumstellar envelopes around evolved stars are very efficient chemistry factories that produce organic and inorganic molecules over very short time scales.  However, our understanding of the interactions between molecular processes, chemical environment,
and local physical conditions is far from complete.  In oxygen- and carbon-rich environments, the reaction routes are clearly different. In oxygen-rich objects, the circumstellar chemistry is believed to be dominated by OH  and H$_2$O molecules, and the chemical complexity of the oxygen-rich environments has been demonstrated by \citet{ziu07}.
In carbon-rich objects, a large fraction of oxygen atoms is locked in CO and C-bearing compounds such as
hydrocarbons and cyanopolyynes are efficiently reprocessed.
Among the carbon-rich objects, the nearby AGB star \object{IRC+10216} has been extensively studied 
 \citep[see, e.g.][and references therein]{cernicharo00,he08}, with  more than 60 molecular species detected.  


This paper is one of a series of studies devoted to the  understanding of circumstellar chemistry of evolved stars. In previous work, we have presented the  spectra of the AGB stars
\object{IRC+10216} \citep{he08} and \object{CIT\,6} \citep{zhang09} and the planetary nebula (PN) \object{NGC\,7027} \citep{zhang08} obtained using the Arizona Radio Observatory (ARO) 12\,m telescope and the Heinrich Hertz Submillimeter Telescope (SMT).  
In the present study, we report a molecular line survey of 
the carbon-rich circumstellar envelope \object{CRL\,3068} using the same 
telescope settings with the aiming at exploring the molecular constituents 
and chemistry processes in an extremely carbon-rich environment.

The carbon-rich AGB star \object{CRL\,3068} 
(\object{AFGL\,3068}, \object{IRAS\,23166+1655}, LL\,Peg) is
a long-period variable with a very long pulsation period of 696 days \citep{bertre95}.
It has a high C/O abundance ratio of 1.38 \citep{winters97} and
has been taken as a prototype of extreme carbon stars \citep{volk92}.
 \citet{jones78} investigated the infrared (IR) spectrum of \object{CRL\,3068}
and detected an SiC absorption feature at 11.3 $\mu$m. The presence of
SiC absorption was further confirmed  by \citet{clement03} based on
the {\it Infrared Space Observatory (ISO}) observations.  The fact that the SiC feature is in self absorption suggests that the circumstellar envelope of CRL 3068 is extremely optically thick.  This is supported by the very low color temperature of the dust emission component \citep[$\sim300\,$\,K, see e.g.][]{lebofsky77,volk92,omont93}.
 \citet{bertre95} found that the dust continuum spectrum of CRL 3068 the wavelength range 1--50\,$\mu$m is consistent with a dust opacity law of $\propto\lambda^{-1.3}$.

Based on a consistent time-dependent hydrodynamical model, \citet{winters97} determined the distance to \object{CRL\,3068} to be 1.2\,kpc, the time averaged outflow velocity as 14.7\,km\,s$^{-1}$, and the average mass loss rate as $1.2\times10^{-4}$\,M$_\sun$yr$^{-1}$.
The distance derived by \citet{winters97}  is close to a more recent value of 1.05\,kpc reported by \citet{yuasa99} and \citet{menzies06}. The derived mass loss rate of \citet{winters97}  is 2--10 times higher than those obtained using CO lines \citep{knapp85,volk93,schoier02,woods03,teyssier06} and that by fitting the dust continuum \citep{volk92}.
The \ion{H}{1} line at 21\,cm was detected by \citet{gerard06}, who estimated that the \ion{H}{1} emission
in \object{CRL\,3068} is extended over a region $\sim8\arcmin$ in
diameter. \citet{mauron06} imaged the extended circumstellar envelope
of \object{CRL\,3068} in the optical and were able to trace the galactic scattered light out to a distance of $\sim40\arcsec$ from the star.
They estimated the expansion timescale of the envelope to be 15,200\,yr. The optical image was characterized by a remarkable spiral pattern traced out to $12\arcsec$ from the center, suggesting that
binary companions might play a role in shaping the circumstellar envelope.  The scenario of binary-induced outflow was further comfirmed by the near-IR adaptive optics images obtained with the Keck~{\sc ii} telescope, which showed that the central region of \object{CRL\,3068} is composed of two components separated by a distance of $0.11\arcsec$ \citep{morris06}. 
\citet{neri98} mapped the CO(1--0) emission, which revealed a compact inner envelope and a surrounding detached bipolar shell.

Molecular lines in \object{CRL\,3068} at millimeter wavelengths have been explored by several authors. Relatively strong CO(1--0) line emission 
has been detected  by \citet{knapp85} and \citet{nyman92}.
HC$_3$N lines were first detected by \citet{jewell84}.
\citet{nguyen88} discovered the HCN(1--0) and H$^{13}$CN(1--0) line
emission.
\citet{sopka89} reported the observations of CO, $^{13}$CO, HCN,
H$^{13}$CN, HC$_3$N, and HNC in this object.
\citet{wannier91} discovered relatively faint emission from C$_3$H$_2$.
Radio observations toward \object{CRL\,3068} in the frequency ranges
between 39--47\,GHz and 85--91GHz were carried out with the Nobeyama 45\,m radio telescope by \citet{fukasaku94}, and the molecular species C$_2$H, HC$_5$N, C$_4$H, and SiS were detected in this object for the first time.
\citet{woods03} reported a molecular line survey made with the SEST 15\,m and Onsala 20\,m telescopes. The molecular species  in \object{CRL\,3068} that were detected positively in their survey observation include CO, $^{13}$CO, CN, HCN, H$^{13}$CN, HNC, C$_2$H, HC$_3$N, C$_3$N, CS, and SiS. 

In this paper, we present spectra of \object{CRL\,3068}, representing the most complete molecular line survey of this carbon-rich envelope so far.  The remaining parts of this paper are organized as follows:
the observations and data reduction are described in Sect.~2; in Sect.~3, we presents the identification and measurements of  detected molecular lines; Sect.~4 gives the column densities and abundances of detected species; in Sect.~5, the implication of our findings on circumstellar chemistry is discussed;
our conclusions are summarized in Sect.~6.

\section{Observations and data reduction}

\subsection{The ARO 12\,m observations}

The spectral line survey at the $\lambda$ 2\,mm window was carried out
during the period from 2007 December to 2008 January with
the ARO 12\,m telescope at Kitt Peak. The 2-mm dual-channel SIS receivers 
were employed, operated in single sideband dual polarization mode.
The image rejection ratio is larger than $18$\,dB on most of the occasions.
The system noise temperatures were typically 250--550\,K.
The spectrometer backends employed are two 256-channel filter banks (FBs) 
with a channel width of 1\,MHz and a millimeter autocorrelator
(MAC) with 3072 channels and 195\,kHz per channel.
The observations were made in beam switching mode with an azimuth beam throw of 2$\arcmin$.  We checked the pointing every 2--3 hours by observations of 
planets.  The pointing accuracy was  better than 15$\arcsec$ at normal
weather conditions.
The scanned frequency range was from  130.0 to 163.9\,GHz with a few gaps.
The half power beam width (HPBW) is about 40$''$ at this frequency range.
Typical on-source integration times ranged from 70--90 minutes for each frequency setting.
For $\lambda$ 2\,mm observations,  we gave priority to the 
frequency ranges covering the strong lines as observed in \object{IRC+10216}
\citep{cernicharo00}. Only a few extremely faint lines in the spectra of
\object{IRC+10216} fall within the gaps of our spectra.
The temperature scale was given in term of $T^*_R$, which was
calibrated by the chopper-wheel method, and 
 corrected for atmospheric attenuation, radiative loss, and rearward and 
forward scattering and spillover.  The main beam brightness
temperature was determined using $T_R=T^*_R/\eta^*_m$, where
$\eta^*_m$ is the corrected beam efficiency ($\sim0.75$).

We used the  CLASS software package in GILDAS \footnote{GILDAS is developed and
distributed by the Observatorie de Grenoble and IRAM.} for the data reduction. After discarding the bad scans, we obtained the co-adding spectra from individual scans.  On some occasions, the MAC spectra might suffer severe effects from the bandpass irregularities, and thus for some frequency coverages the signal-to-noise ratios might be relatively low due to the exclusion 
of these scans.
The baseline was determined by low-order polynomial fits to the line-free regions of the spectra.
In order to improve the signal-to-noise ratio, the spectra were smoothed and rebinned to a frequency resolution of 1\,MHz, corresponding a velocity resolution of $\sim2$\,km\,s$^{-1}$.
The MAC spectra are presented in Figure~\ref{spe_12m}. The typical rms noise temperature is less than 15\,mK in main beam temperature unit.

\subsection{The SMT observations}

The spectral line survey at the $\lambda$ 1.3\,mm window was made using the SMT 
10\,m telescope
located on Mt. Graham in 2008 January. For the observations, we utilized the ALMA Band 6 (211--275\,GHz) sideband-separating
SIS mixing-preamp, yielding a typical image rejection ratio of $>20$\,dB and a system noise temperature of $\sim150$\,K.  The 2048-channel acousto-optical spectrometer (AOS) and the 1024-channel Forbes Filterbanks (FFBs) were used simultaneously and were configured to provide a channel width of 500\,kHz and 1\,MHz, respectively.
A beam switching mode was applied with a separation between the ON and OFF beams of 2$\arcmin$ in azimuth.  The pointing was checked every 2 hours on nearby planets and was found to be accurate to within 10$\arcsec$.
The complete frequency range from 219.5 to 267.5\,GHz was covered, with typical on-source integration time of 30 minutes at each frequency band. 
The beam size at this frequency range is about 30$''$.
The data were calibrated to the  $T^*_A$ temperature scale,
which was corrected for atmospheric attenuation. The main beam temperature was derived using $T_R=T^*_A/\eta_{mb}$, where $\eta_{mb}$ is the main beam efficiency ($\sim0.7$).

The procedure of data reduction was same as that for the 12\,m data. Since there is no gap in the $\lambda$ 1.3\,mm survey region, all the spectral bands were merged together.  Figure~\ref{spe_smt} gives the FFB spectra, 
which have been smoothed and rebinned by a factor of 3, yielding a frequency resolution of 3\,MHz, corresponding a  velocity resolution of 3.7\,km\,s$^{-1}$, and a typical rms temperature of 5\,mK in main beam temperature unit.

\section{Observational results}

\subsection{The overall survey}

The spectra together with line identifications are shown in Figures~\ref{spe_12m} and \ref{spe_smt}. Note that some features in these figures are not real but caused by bandpass irregularities or bad channels.
A total of 75 distinct emission features (including a few uncertain ones) were detected in our survey 
observations. The most intense line detected in our survey is the CO (2--1) 
transition, followed by the HCN (3--2) transition. Apart from these two transitions, the others have integrated intensities of less than 10\,K\,km\,s$^{-1}$.
The molecular lines detected in the high frequency window are
a factor of about two more than those detected in the low frequency
window.  This is partially due to the fact that the ARO 12\,m telescope has a larger beam size than the SMT and thus the 12\,m data are more likely to be affected by the beam dilution.
The molecular line databases that are used for line identifications include the archives of molecular line frequencies derived from the theoretical calculations of the JPL catalog \citep{pickeet98}\footnote{http://spec.jpl.nasa.gov.} and and the Cologne database for molecular spectroscopy \citep[CDMS,][]{muller01,muller05}\footnote{http://www.ph1.uni-koeln.de/vorhersagen/.},
and those from previous observations of other sources \citep[e.g.] [and the NIST Recommended Rest Frequencies for Observed Interstellar Molecular Microwave Transitions\footnote{http://physics.nist.gov/cgi-bin/micro/table5/start.pl}]{cernicharo00,pardo07,he08,zhang09}.
We identified 11 main molecular species, including
CO, CS, C$_2$H, C$_4$H, $c$-C$_3$H, CH$_3$CN, CN, HCN, SiC$_2$, SiO, and SiS,
and 12 rare isotopologues, including $^{13}$CO, C$^{17}$O, C$^{18}$O, C$^{33}$S, C$^{34}$S, $^{13}$CS, HC$^{15}$N, H$^{13}$CN, HN$^{13}$C, HC$_3$N, $^{29}$SiS, and $^{30}$SiS.
Only three fairly weak features remain unidentified.
Several transitions are new detection in \object{CRL\,3068}.
A full list of line identifications and measurements is presented
in Table~\ref{line}. Columns~1--3 give the identified species, 
transitions, and frequencies, respectively. The
rms noise levels are given in column~4.  Column~5 and 6
lists the peak and integrated intensities. Column~7 gives
the line widths ($FWHM$) obtained by fitting Gaussian line profiles.
For blended lines, Table~\ref{line} gives the intensities and widths
of the combined features.

Figure~\ref{cum} shows the cumulative number of detected lines exceeding a given integrated intensity. For comparison, the figure also plots the results for the spectra of \object{IRC+10216} and \object{CIT\,6} which were obtained in the same observation project \citep{he08,zhang09}. 
The shapes of these curves in Figure~\ref{cum} are similar with each other although that of \object{CRL\,3068} seems to be slightly steeper than the other two.  This suggests that the differences between the three objects can be mostly a distance effect, and the detection of a large number of lines in IRC+10216 does not imply that IRC+10216 is unique or richer in molecular content than the other two.

\subsection{Individual molecules}
The details of the molecules detection in \object{CRL\,3068}
are described below. Here we compare these detections with
the spectra of \object{IRC+10216} and \object{CIT\,6},
which are respectively from \citet{he08} and \citet{zhang09} if no statement
is issued.

\subsubsection{CO}

The $J=2$--1 transitions of CO and its isotopologues  $^{13}$CO, C$^{17}$O, and C$^{18}$O were detected by the SMT.  The CO (2--1) transition is the strongest line detected in this survey, with an integrated intensity of 35.9\,K\,km\,s$^{-1}$. This line has been observed by \citet{volk93} using the 15\,m JCMT.
They obtained an integrated intensity of $\int T_A^*$d$v=40.2$\,K\,km\,s$^{-1}$ (i.e. $\int T_{\rm R}$d$v=57.4$\,K\,km\,s$^{-1}$), slightly larger than our result. A higher integrated intensity of the CO (2--1) line was detected by \citet{groenewegen96} using the IRAM telescope.
However, considering the larger beam size of the SMT compared to those of JCMT and IRAM, these intensity discrepancies are insignificant. 
The $^{13}$CO (2--1) line has also been detected by \citet{groenewegen96} and \citet{bujarrabal94}. Our result yields an intensity ratio of $I$(${^{13}}$CO 2--1)/$I$(${^{12}}$CO 2--1)$=0.14$, larger than the values of 0.10 and 0.07 found in \object{IRC+10216} and \object{CIT\,6}.
This reflects differnt $^{13}$C/$^{12}$C isotopic ratios and/or
different optical depth effects on the three objects.  The faint C$^{17}$O and C$^{18}$O lines are only marginally above the detection limit. The two rare isotopologues should be new detections for this object.  The CO line shows a narrower profile than the isotopic lines which can be explained by the fact that the former is optically thick.

\subsubsection{CS}

The $J=3$--2 and $J=5$--4 transitions of CS were detected by
the ARO 12\,m and the SMT, respectively. The two CS lines toward
this source have also been detected by \citet{bujarrabal94} using the 
IRAM telescope. In the $\lambda$ 1.3\,mm window,
we for the first time detected the $J=5$--4 isotopic transitions of
C$^{33}$S, C$^{34}$S, and $^{13}$CS,
among which the C$^{33}$S (5--4) line emission is relatively faint and its 
measurement is uncertain. The  $J=3$--2  isotopic transitions lie in the 
$\lambda$ 2\,mm region, but are below our detection limit.
The two main lines have a similar width but are narrower than the isotopic lines, suggesting that the main lines are optically thick.
We obtained the intensity ratios of $I$(${^{13}}$CS 5--4)/$I$(${^{12}}$CS 5--4)$=0.09$ and  $I$(C${^{34}}$S 5--4)/$I$(C${^{32}}$S 5--4)$=0.13$, larger than the corresponding values of 0.03 and 0.07 in \object{IRC+10216}. The trend
is the same as that found for the $I$(${^{13}}$CO 2--1)/$I$(${^{12}}$CO 2--1)
ratio.


\subsubsection{C$_2$H}

\citet{fukasaku94} detected C$_2$H  in this object through the  $N=1$--0 
transition.  Their detection was further confirmed by \citet{woods03}.
The $N=3$--2 transition is split in six hyperfine-structure 
lines grouped in three fine-structure groups, which lie in our
survey range. The two strong main components, 3$_{7/2}$--2$_{5/2}$ and
3$_{5/2}$--2$_{3/2}$, are clearly detected in the 2\,mm window.
The 3$_{7/2}$--2$_{5/2}$ transition is blended with a considerably 
weaker SiC$_2$ line.  The faint 3$_{5/2}$--2$_{5/2}$ transition is 
overwhelmed by noise. We obtained the intensity ratio of
$I$(C$_2$H 3$_{7/2}$--2$_{5/2}$)/$I$(C$_2$H 3$_{5/2}$--2$_{3/2}$)$=1.7$,
in good agreement with the values of 1.4 and 1.5 found in
\object{IRC+10216} and \object{CIT\,6}, respectively.

\subsubsection{C$_4$H}

The $N=10$--9 transition of C$_4$H in this object has been
detected by \citet{fukasaku94}.  
Due to fine structure interaction, every rotational transition of C$_4$H is split into two components, which have  similar intensities.
A total of 12 favorable lines of C$_4$H lie in our survey range. The eight C$_4$H lines in the $\lambda$ 2\,mm window were clearly detected. The four lines in the $\lambda$ 1.3\,mm window are too faint to obtain reliable  measurements. In Table \ref{line}, we only give an uncertain detection of the C$_4$H (24--23 a,b) lines, which is the strongest transition  of  C$_4$H in the $\lambda$ 1.3\,mm window.
We found that these high-$J$ C$_4$H  lines are at least 3 times fainter than those in the $\lambda$ 2\,mm  window. 
In \object{IRC+10216}, however, these C$_4$H transitions have a similar strength \citep[see Table~9 of][]{he08}. This cannot be completely attributed to different beam dilution effects in the two objects, and
can be a reflection of  the different physical conditions of the two carbon envelopes.

\subsubsection{$c$-C$_3$H}

The cyclopropynylidyne radical ($c$-C$_3$H) has not been detected 
in \object{CRL\,3068} before this work. 
Here we report the first tentative detection of this species.
Six favorable transitions of $c$-C$_3$H lie in our survey region.
Only the strongest two transitions are discovered in our spectra
with relatively  weak strengths. $c$-C$_3$H has been detected in \object{IRC+10216}, whereas there is no evidence for the corresponding lines of $c$-C$_3$H in \object{CIT\,6}.

\subsubsection{CH$_3$CN}

There was no report of the detection of CH$_3$CN  in this object before
this work.  \citet{woods03} estimated the intensity upper limit of the 
(6$_1$--5$_1$)
transition. Here we present the first detection of CH$_3$CN  in this object.
There are a total of 14 favorable CH$_3$CN transitions in our survey region.  
The strongest two fine-structure groups are clearly detected
in the $\lambda$ 1.3\,mm window, but the others are below our detection limit.
This species was also detected in \object{IRC+10216} and \object{CIT\,6}.

\subsubsection{CN}

The cyanogen radical (CN)  is one of the most abundant molecules in 
envelopes of carbon rich stars.
CN in \object{CRL\,3068} has been discovered by \citet{woods03} and \citet{bachiller97} in the  $N=1$--0 and $N=2$--1 transitions.
There are three $N=2$--1 fine-structure groups of CN  present in our survey region. They are split into total of 18 hyperfine structure components.
All the three groups were apparently observed in the $\lambda$ 2\,mm window.
The weakest one around 226.33\,GHz should be a new detection.  Our observations suggest the intensity ratio of the  fine-structure groups around 226.87\,GHz and 226.69\,GHz to be 1.6, 
in excellent agreement the value of 1.4 obtained by \citet{bachiller97}.  The value is also consistent with those found in \object{IRC+10216} and \object{CIT\,6}.

\subsubsection{HCN}

The $J=3$--2 transitions of HCN and its isotopologues HC$^{15}$N and H$^{13}$CN were clearly
detected with  well-defined profiles in the $\lambda$ 2\,mm window.  The HCN (3--2) transition is the second strongest line after the CO (3--2) transition in our survey.  A few vibrationally excited lines of HCN have been detected in \object{IRC+10216} and \object{CIT\,6}. The strongest two are detected in the spectra of \object{CRL\,3068}. To our  knowledge, this is  the first detection of vibrationally excited lines in this object. We discovered the isotopologue HC$^{15}$N for the first time. The intensity ratio of $I$(HC$^{15}$N 3--2)/$I$(H$^{13}$CN 3--2) is 0.027, which is slightly higher than the value of 0.016 obtained in \object{IRC+10216}. 
We also obtained the intensity ratio of $I$(H$^{13}$CN 3--2)/$I$(H$^{12}$CN 3--2)$=0.60$, larger than the values of 0.16 and 0.41 found in \object{IRC+10216} and \object{CIT\,6}.  The larger isotopologue-to-main line ratios in \object{CRL\,3068} are also found for CO and CS (see above).  Consequently, we infer that the lines in this extreme carbon star is more optically thick than those in \object{IRC+10216} and \object{CIT\,6}.

\subsubsection{HNC}

HNC has been detected in  \object{CRL\,3068} \citep{sopka89,fukasaku94,woods03}. This species can be produced through a similar way to HCN. HNC can also be enhanced through ion-molecular reactions or be transferred into HCN in high temperature.  An abundance comparison of HNC with HCN can provide some insight into circumstellar chemistry.
Although no HNC line lies in our survey region, we conclusively detected its isotopologue HN$^{13}$C in the $J=3$--2 transition for the first time. Our observations suggest the intensity ratio 
$I$(H$^{13}$CN 3--2)/$I$(HN$^{13}$C 3--2) to be 19.3, about an order of magnitude lower than that in \object{IRC+10216}.  This  implies that HNC has been enhanced in \object{CRL\,3068}.  HN$^{13}$C was not detected in  \object{CIT\,6}.

\subsubsection{HC$_3$N}

Cyanoacetylene (HC$_3$N) has been commonly detected in carbon-rich circumstellar envelopes. The presence of HC$_3$N in \object{CRL\,3068} has been confirmed \citep{jewell84, sopka89, bujarrabal94, fukasaku94,woods03}.  Nine  HC$_3$N transitions from $J=15$--14 to $J=29$--28 lie in the frequency range of our survey.
All of them were detected in our spectra.  Most of these high-$J$ transitions are new detections.

\subsubsection{SiC$_2$}

SiC$_2$ is the species having the most emission lines in our survey.  Including possible detections and blended features, a total of 23 SiC$_2$ transitions were detected. These SiC$_2$ lines are relatively weak with peak brightness temperatures less than 40\,mK.  We cannot find  previous papers reporting on the detections of SiC$_2$ in \object{CRL\,3068}.  \citet{woods03} failed to discover SiC$_2$ in this object and estimated an intensity upper limit of the SiC$_2$ (5$_{0,5}$--4$_{0,4}$) transition.

\subsubsection{SiO}

SiO in \object{CRL\,3068} has been detected by \citet{bujarrabal94} in the $J=2$--1 and $J=3$--2 transitions.  In our survey the $J=3$--2 and $J=6$--5 transitions were detected.  The SiO (6--5) transition was also detected in our spectra of \object{IRC+10216} and \object{CIT\,6}. We found that the relative intensity of the SiO transition in \object{CRL\,3068} is much lower than those in \object{IRC+10216} and \object{CIT\,6}. We obtained the intensity ratio of  $I$(SiO 3--2/$I$($^{13}$CO 3--2) in \object{CRL\,3068} is 0.1, an order of magnitude lower than the values of 1.1 and 1.0 found
in \object{IRC+10216} and \object{CIT\,6}. Its isotopic transitions $^{29}$SiO and  $^{30}$SiO  (3--2)
lie in the surveyed range but are overwhelmed by noise.

\subsubsection{SiS}

The $J=6$--5 and $J=5$--4 transitions of SiS in \object{CRL\,3068} have previously been observed by \citet{bujarrabal94}, \citet{fukasaku94}, and \citet{woods03}. In this survey we conclusively detected four SiS transitions with higher $J$ (8--7, 9--8, 13--12, and 14--13) transitions.  We also report the first possible detection of the isotopologues $^{29}$SiS and $^{30}$SiS.

\subsubsection{Unidentified lines}

Three weak lines detected at 3--8$\sigma$ noise levels
remain unidentified.  The 264067\,MHz line probably corresponds to the
U264072 line detected in \object{IRC+10216} and
the other two U lines have no corresponding  detection in 
\object{IRC+10216} and \object{CIT\,6}.
We also searched the NIST frequency table and
found that the two features at 245982\,MHz and 255108\,MHz seem
to be associated the U lines at 245993\,MHz and 255158\,MHz
detected in Sgr B2(N) \citep{num98}.

\subsection{Line profiles}

Since the circumstellar envelopes of AGB stars are expanding, molecular line profiles can provide significant insight into the nature of the lines as well as the kinematic structure of the envelopes.
The lines arising from a spherical envelope might have
four characteristic profile shapes \citep[see e.g.,][]{habing03, kwo06}: (a) 
a flat-topped line profile resulting from optically thin, spatially unresolved 
emission; (b) a parabolic line profile resulting from optically thick,
spatially unresolved emission; (c) a double-peaked line profile resulting from 
optically thin, spatially resolved emission; (d) a smoothed parabolic
(or seemingly flat-topped)  line profile results from optically thick, spatially resolved emission. Our observations show that most of the detected lines are consistent of they being unresolved.
The profiles of the CO, CS, and HCN emission lines appear to be parabolic, indicating that they are optically thick. The $^{13}$CO line show a
rectangular profile, suggesting that it is likely optically thin.
Some of the lines, such as vibrationally excited lines of HCN, exhibit fairly
narrow profiles, probably because they are excited in a compact region close 
to the central star. The different widths for different emission lines 
seem to suggest that the expansion velocity of the envelope is not constant.

\subsection{Nondetection}

Compared with the spectra of \object{IRC+10216}, the  
nondetected molecular species in our survey region include
AlCl, AlF, MgNC, NaCN, NaCl, PN, C$_2$S, C$_3$H, $c$-C$_3$H$_2$,
C$_3$N, C$_3$S,  $l$-C$_4$H$_2$, H$_2$CO, SiC, and SiN. 
All the emission lines from these species are weak in \object{IRC+10216}
and are estimated to have an integrated-intensity upper-limit of 
0.2\,K\,km\,s$^{-1}$ in \object{CRL\,3068}. If the line intensities relative to the $^{13}$CO (2--1) transition in \object{CRL\,3068} are the same as those in \object{IRC+10216}, all these lines are well below our detection limit. Therefore, we cannot draw the conclusion that the molecular species in  
\object{CRL\,3068} are less abundant than in \object{IRC+10216}.

\section{Column densities and abundances}

For comparison, we have applied the same approach as used in our previous papers \citep{zhang08, zhang09}
to calculate the column densities and abundances of the molecules detected in our spectra. This will minimize the effects of systematic errors when we compare the results in different objects.

The excitation temperatures  ($T_{\rm ex}$) and column densities ($N$) were deduced using the standard rotational-diagram analysis. Under the assumption that
the lines are optically thin, the level populations are
in local thermal equilibrium (LTE), and $T_{ex}>>T_{bg}$, where $T_{bg}$ is the cosmic background
radiation temperature (2.7\,K), we have the relation,
\begin{equation}
\ln \frac{N_u}{g_u}=\ln\frac{3k\int T_s dv}{8\pi^3\nu S\mu^2}=
\ln\frac{N}{Q(T_{\rm ex})}-\frac{E_u}{kT_{\rm ex}},
\end{equation}
where $N_u$, $g_u$, and  $E_u$ is
the population, degeneracy, and excitation energy of the upper level,
$\int T_s dv$ is the integration  of the source brightness temperature
over the velocity, $S$ is the line strength,
$\mu$ is the dipole moment, $\nu$ is the line frequency,
 $Q$ is the rotational partition function.
If several transitions arising from levels covering a wide energy range
are observed, $T_{\rm ex}$ and $N$ can be determined using
a straight-line fit to ${N_u}/{g_u}$ versus ${E_u}/{kT_{\rm ex}}$.
Departure from the linear relation suggests different excitation
mechanisms or misidentification.
For the calculations, we have corrected the effect of beam dilution through
 $T_s=T_R(\theta^2_b+\theta^2_s)/\theta^2_s$, where 
$\theta_b$ ($\sim40${\arcsec} and 30{\arcsec}
for the ARO 12\,m and the SMT respectively)
is the antenna full beam at half-power, and $\theta_s$  is the source diameter.
Different transitions may have different $\theta_s$ and it is hard to determine $\theta_s$ for each molecule. According to the CO observations reported by \citet{neri98} and \citet{teyssier06}, we assumed 
$\theta_s=22\arcsec\pm8\arcsec$. The assumed $\theta_s$ value is similar to that used by \citet{fukasaku94} and will introduce a $\sim70\%$ uncertainty in the derived column densities.  
Four molecular species, SiC$_2$, SiS, HC$_3$N, and C$_4$H have adequate numbers of detected transitions covering a wide range of excitation energy and the rotational diagrams of these species are given in Figure~\ref{dia}. 
Although in general the transitions of these species are well fitted by straight lines, we note that the rotational diagram of SiC$_2$ indicates a higher $T_{\rm ex}$ for high-$J$ transitions compared to the low-$J$ ones. This is similar to what is  found in \object{IRC+10216} \citep{avery92, he08}, suggesting
that excitation by infrared radiation may be important for this molecule in AGB envelopes.
For the species for which the rotational diagrams are unavailable,  an average $T_{\rm ex}$ ($\sim53$\,K) was assumed to calculate their column densities. 

The molecular abundances respect to molecular hydrogen ($f_{\rm X}$) were calculated using the
expression by \citet{olofsson96},
\begin{equation}\label{abundance}
f_{\rm X}=1.7\times10^{-28}\frac{v_e\theta_bD}{\dot{M}_{{\rm H}_2}}
\frac{Q(T_{\rm ex})\nu_{ul}^2}{g_uA_{ul}}
\frac{e^{E_l/kT_{\rm ex}}\int T_Rdv}{\int^{x_e}_{x_i}e^{-4x^2\ln2}dx},
\end{equation}
where $\int T_Rdv$ is given in K\,km\,s$^{-1}$,
the full half power beam width $\theta_b$ is in arc\,sec,
the expansion velocity $v_e$ is in km\,s$^{-1}$, $D$
is the distance in pc, $\dot{M}_{{\rm H}_2}$ is
the mass loss rate in $M_{\sun}\,{\rm yr}^{-1}$, $\nu_{ul}$  the
line frequency in GHz, $g_u$ is the statistical weight of
the upper level, $A_{ul}$ is the Einstein coefficient for the transition,
$E_l$ is the energy of the lower level, and $x_{i,e}=R_{i,e}/(\theta_bD)$
with $R_i$ and $R_e$ the inner radius and outer radius of the shell.
To aid the reader, Appendix~A provides full deduction of this formula.
 For the abundance calculation, we first determined the mass loss rate 
 using the CO (2--1)  line. Using the formula of \citet{winters02} and assuming $f_{\rm CO}=1\times10^{-3}$ and a distance of 1.05\,kpc, we obtained $\dot{M}_{\rm H_2}\approx3\times10^{-5}$\,$M_{\sun}\,{\rm yr}^{-1}$,  in good agreement with the recent result of \citet{schoier02} and \citet{woods03} but a factor of two lower than that of \citet{teyssier06}. 
Whenever available, the $R_i$ and $R_e$ values given by \citet{woods03}
were used. The emission regions of isotopologues were assumed to have the same 
inner and outer radii with those of their main species.
The $T_{\rm ex}$ values derived from rotation diagram analysis are
adopted. For species where more than one line is detected,
we find that the spreads of the abundances derived from different
transitions are very small ($<10\%$), and determine the average values
weighted by line intensity.
Although the adopted $T_{\rm ex}$ value is significantly larger
than the average excitation temperature derived by \citet{woods03}
for a sample of carbon stars and the excitation temperature may not be 
homogeneous within the envelope, the abundances are essentially 
insensitive to $T_{\rm ex}$  in the case of $E_l<<kT_{\rm ex}$.

The derived values for  $T_{\rm ex}$, $N$, and $f_{\rm X}$ are given in Table.~\ref{abundance} . Our calculations are based on an optically thin assumption. Therefore, for the optically
thick emission, the $N$ and $f_{\rm X}$ given in Table~\ref{abundance}
should be treated as lower limits. Combining various error sources
(uncertainties of line intensity, excitation temperature, distance, mass-loss 
rate, etc.), we crudely estimate that the errors of the absolute column 
densities and abundances amount to a factor of $\sim5$. Note that the estimated
errors are for the case that the basic assumptions of the method, i.e.
optically thin emission, are met. If the extent of a given species
is significantly different from that of CO, its column density may 
bear a larger error than estimated here. It is impossible to accurately
determine the extent of each species without interferometric observations.
In most of the cases, the self-shielding molecule CO is more extended than 
other molecules \citep[see, e.g.,][]{woods03}. If the source size is 
overestimated by a factor of two, the resultant column density will
been underestimated by a factor of four. Accordingly, it is possible that
the actual errors could rise above a factor of ten. Nevertheless, the column 
density ratios and the abundance ratios should be much more reliable.

Table~\ref{abundance} also gives a comparison of our results with those derived by \citet{fukasaku94} and \citet{woods03}. While more molecular species are detected in our survey, for the molecules  that are commonly detected  (but no necessarily from the same transitions), Table~\ref{abundance} shows that
the $N$ and $f_{\rm X}$  values are in agreement within one order of magnitude. Such discrepancies are understandable considering 
different instruments and methods used for the observations and
abundance analysis.

\section{Discussion}

\subsection{Isotopic species}

Theory predicts that during the AGB phase the nucleosynthesized products in 
the inner shell are dredged up to the stellar surface, resulting in changing 
the isotopic ratios of circumstellar envelopes \citep[see][for a recent review]{busso06}. 
Extensive observations have shown that the isotopic composition in carbon-rich stars
is remarkably non-solar \citep[e.g.][]{cernicharo00}, providing substantial
tests for nucleosynthesis theory of AGB stars. Such non-solar isotopic ratios also make possible the identification of stellar solid-state materials in the Solar System \citep{zin98}.

In Table~\ref{isoto}, 
we give the isotopic ratios (or their lower limits) of carbon, oxygen, nitrogen,
sulfur, and silicon in the extreme carbon star, \object{CRL\,3068}. This table
also lists the isotopic ratios for \object{CIT\,6}, \object{IRC+10216}, and the Sun.
A comparison of isotopic ratios in these objects allow us to study 
the nucleosynthesis and dredge-up processes in diverse 
 physical conditions.

\subsubsection{Carbon}

The $^{12}$C/$^{13}$C isotopic ratio is a good indicator of interior 
nucleosynthesis processing. The dredge-up processes can greatly enhance
$^{13}$C in the surface. Moreover, the cool bottom processing and
hot bottom burning can further decrease the $^{12}$C/$^{13}$C ratio
\citep[e.g.][]{boothroyd99,frost98}.

We detected three $^{13}$C-bearing species, $^{13}$CO, $^{13}$CS, and 
H$^{13}$CN. However, the CO/$^{13}$CO, CS/$^{13}$CS, and HCN/H$^{13}$CN
ratios only give lower limits of the $^{12}$C/$^{13}$C because the
main lines appear to be optically thick. On the other hand, if
the $^{32}$S/$^{34}$S isotopic ratio were known, we could use
the optical thin  $^{13}$CS and  C$^{34}$S lines to obtain the
 $^{12}$C/$^{13}$C ratio. Based the fact that the nucleosynthesis in AGB 
stars is expected not to affect the abundances of sulfur isotopologues
and that \citet{cernicharo00} has found  the sulfur isotopic ratio
in \object{IRC+10216} to be close to solar, we can reasonably assume that
the $^{32}$S/$^{34}$S ratio in \object{CRL\,3068} is solar (22.5).
From this assumption, we obtained a $^{12}$C/$^{13}$C ratio of 29.7.
Using a full radiative transfer analysis of $^{12}$CO and $^{13}$CO
line emission, \citet{woods03} derived the $^{12}$C/$^{13}$C ratio
of 30, lending strong support to our result.
The $^{12}$C/$^{13}$C ratio in \object{CRL\,3068}
is significantly lower  than the solar value,
suggesting the presence of extra mixing processes.

As a consequence of dredge-up processes, the C/O abundance ratio in
circumstellar envelopes decreases along with the $^{12}$C/$^{13}$C ratio.
As shown in  Table~\ref{isoto}, the $^{12}$C/$^{13}$C ratio
in \object{CRL\,3068} is about a factor of 1.5 lower than those
found in \object{CIT\,6} and  \object{IRC+10216}, suggesting that
\object{CRL\,3068} has a higher C/O ratio.   Our observations, therefore, 
provide an opportunity to study 
circumstellar chemistry in extremely carbon rich environment.

\subsubsection{Nitrogen}

As a star ascends the AGB, the $^{14}$N-rich and  $^{15}$N-poor material processed by the CNO cycle in the intershell is dredged up to the stellar surface, resulting a high $^{14}$N/$^{15}$N abundance ratio in the envelope. \citet{wannier91} determined the lower limits of $^{14}$N/$^{15}$N in a sample of carbon stars and found that in most of the cases they were larger than the terrestrial ratio.

The nitrogen isotopologue, $^{15}$N, has been detected through
a faint emission line from HC$^{15}$N. As its main line is optically
thick, the HC$^{14}$N/HC$^{15}$N abundance ratio gives a
lower limit of the isotopic ratio of 45. The optically thin species
H$^{13}$C$^{14}$N and H$^{12}$C$^{15}$N may also be applied to derive the 
$^{14}$N/$^{15}$N  ratio. Assuming a $^{12}$C/$^{13}$C ratio
of 29.7 (see above), we obtained the $^{14}$N/$^{15}$N ratio of 1099,
a factor of 4 higher than the solar value.
 \citet{wannier91} obtained a lower limit of $^{14}$N/$^{15}$N$>500$
for \object{CRL\,3068}, in good agreement with our result.
  The deduced nitrogen
isotopic ratio is in accord with the predictions of stellar models.

\subsubsection{Oxygen}

Nucleosynthesis models predict enhancement of $^{17}$O in the circumstellar envelopes around carbon stars \citep{busso06}.  Previous observations have shown that the C$^{17}$O/C$^{18}$O ratios in carbon-rich envelopes are significantly higher than the terrestrial and interstellar values \citep{wannier87,kahane92}.

The weak C$^{18}$O and C$^{17}$O (2--1) transitions were detected in our survey, allowing us to determine the oxygen isotopic ratios in \object{CRL\,3068}. The abundance ratios of CO and its isotopologues gives lower limits of isotopic ratios, as can be seen in  Table~\ref{isoto}. Under the assumption of $^{12}$C/$^{13}$C$=29.7$, we used the optically thin species $^{13}$CO, C$^{18}$O, and C$^{17}$O to obtain the $^{16}$O/$^{17}$O, $^{16}$O/$^{18}$O, and $^{17}$O/$^{18}$O ratios of 668, 472, and 0.7, respectively. The $^{16}$O/$^{17}$O and $^{17}$O/$^{18}$O ratios bear a large uncertainty, but clearly point  to an enhancement of $^{17}$O.  The $^{16}$O/$^{18}$O ratio is comparable with the solar value.  These isotopic ratios are well within the ranges reported by
\citet{kahane92} for a sample of carbon-rich envelopes.

\subsubsection{Sulfur and silicon}

The isotopic compositions of sulfur and silicon are hardly
affected by nucleosynthesis of AGB stars although they
might be slightly changed by neutron capture. This is
consistent with the observations that the S and Si
isotopic ratios in \object{IRC+10216} and \object{CIT\,6}
are close to the solar values \citep{cernicharo00,zhang09}.

We detected the S and Si isotopologues  $^{33}$CS, $^{34}$CS, 
$^{29}$SiS, and $^{30}$SiS in \object{CRL\,3068}. The main line
CS (3--2) is optically thick, while the few SiS transitions
are likely to be optical thin. Consequently, we derived 
the $^{33}$S/$^{34}$S ,  $^{29}$Si/$^{30}$Si,  $^{28}$Si/$^{30}$Si,
and  $^{28}$Si/$^{29}$Si ratios,  and the lower limit
of the $^{32}$S/$^{34}$S ratio, as listed in Table~\ref{isoto}.
In \object{IRC+10216} and \object{CIT\,6}, the Si isotopic ratios
were also obtained from the SiO and SiC$_2$ isotopologues 
\citep{he08,zhang09}, which, however, were not detected in 
the spectra of \object{CRL\,3068}. 
We found that these S and Si isotopic ratios are compatible with the solar 
values, and there is no significant deviation between the S and Si isotopic
 ratios
in \object{CRL\,3068} and those in \object{IRC+10216} and \object{CIT\,6}.

\subsection{Remarks on the comparison between CRL\,3068 and IRC+10216}

\object{CRL\,3068} and \object{IRC+10216} are the prototypes of extreme carbon stars, which are highly evolved heavily obscured by thick dust shells \citep{volk92}.  Nevertheless, there exist some differences between the two objects. \object{CRL\,3068} has a colder IR color temperature than \object{IRC+10216} 
\citep[e.g. see][]{omont93}. 
Based on the ({\it ISO}) observations, \citet{justtanont98} have determined the color temperatures of \object{CRL\,3068} and \object{IRC+10216} to be of 320 and  540\,K, respectively.
The {\it ISO} spectra show that the SiC 11.3 $\mu$m feature exhibits a weak flat-topped emission profile in  \object{IRC+10216} while in \object{CRL\,3068} it is in absorption.  This suggests either that \object{CRL\,3068} has a thicker dust envelope, or the abundance of the SiC grains is higher.  Both the color temperature comparison and the SiC feature are consistent with the suggestion that  \object{CRL\,3068} is in an even more evolved state than IRC+10216.

In Figure~\ref{linecomp}, we compare the integrated intensity ratios of the lines detected in  \object{CRL\,3068} and \object{IRC+10216}.
The average line ratio between the two objects is 0.06 with a standard deviation of 0.07.
While there is a large scatter of ratios between species, for individual species the ratios are quite  consistent. For example, the intensity ratios of SiC$_2$ lines have an average of 0.022 with a standard deviation of 0.005.
Note that in this analysis, we have not taken into account the beam dilution effect. 
Using the same method in \citet{zhang09} and assuming that \object{CRL\,3068} and \object{IRC+10216} have diameters of 22$''$ and 30$''$ respectively, we estimate that these intensity ratios would increase by a factor of 1.6 and 1.4 for the ARO 12\,m and the SMT data, respectively if the beam dilution effect is corrected for\footnote{According to the modelling by \citet{woods03}, \object{CRL\,3068} has almost same CO envelope size as \object{IRC+10216}. If this
is the case, the intensity ratios are not affected by the beam dilution effect.}.  \object{CRL\,3068}  is about 8--10 times more distant than  \object{IRC+10216}, and thus might intrinsically  have a stronger average line intensity when the distance factor is taken into account.

Referring to Figure~\ref{linecomp}, if taking the average line intensity ratio as a reference, one can find that the emission from
SiO, SiC$_2$, SiS, CS, HCN, and their isotopologues seems to be depleted in  \object{CRL\,3068} compared to those in  \object{IRC+10216},
while  HN$^{13}$C, $c$-C$_3$H, HC$_3$N, and CH$_3$CN might be enhanced. 
The situation for other species is ambiguous. Although the classification is not very strict, it reflects that \object{CRL\,3068} has different chemical composition, excitation conditions, and optical depths with \object{IRC+10216}. 
A detailed discussion of the implication for
circumstellar chemistry will be given in the following 
section.

Our results are in contrast to that presented in \citet{woods03}, who 
found that for a sample of carbin-rich circumstellar
envelops, the abundance discrepancies are typically less than a factor
of five and the two objects, \object{IRC+10216} and \object{CRL\,3068}, 
are chemically similar. However, in \object{CRL\,3068}, \citet{woods03} did not detect the 
species having the largest abundance discrepancies found here (e.g. SiO and
$c$-C$_3$H). Recent studies of SiO and SiS in samples of carbon stars have 
been presented by \citet{schoier06} and \citet{schoier07}, who derminated the
molecular abundances through a detailed radiative transfer modelling of
the SiO and SiS lines observed by them and some other authors. Their results
suggest that the fractional abundances of SiO and SiS in carbon stars
are in good agreement with each other. If the SiO and SiS abundances
in \object{CRL\,3068} are similar to those in \object{IRC+10216}, our observations
would suggest a significant enhancement of carbon chain molecules in \object{CRL\,3068},
as shown in Figure~\ref{linecomp}. Nevertheless, we cannot completely
rule out the possibility that the Si-bearing molecules in \object{CRL\,3068} are actually depleted.

\subsection{Implication for circumstellar chemistry}

The spectra of \object{CRL\,3068} are characterized by the wealth of C-bearing molecules. Among the newly detected species, only  two are non-C-bearing molecules  (SiO and SiS). From  Figure~\ref{linecomp}, we can
see that  \object{CRL\,3068} has generally higher intensity ratios of C-bearing molecular lines to non-C-bearing ones compared to \object{IRC+10216}.  This seems to lend support to the idea that  \object{CRL\,3068} is more evolved and more carbon rich than  \object{IRC+10216}.	

It is intriguing that according to  Figure~\ref{linecomp} \object{CRL\,3068} 
shows enhancement of HN$^{13}$C and depletion of H$^{13}$CN compared to 
\object{IRC+10216}.  Assuming that carbon isotope fractionation is insignificant for the two species and  the HN$^{13}$C/HNC abundance ratio is equal to the H$^{13}$CN/HCN abundance ratio, we can infer that  \object{CRL\,3068}  has a higher HNC/HCN abundance ratio than \object{IRC+10216}. 
The  HN$^{13}$C (3--2)/H$^{13}$CN (3--2) integrated intensity ratio in \object{CRL\,3068} is 0.05, about
an factor of 13 larger than the value in \object{IRC+10216} \citep[see][]{he08}. HCN and HNC can be produced through a similar way, i.e. dissociative recombination of HCNH$^+$, and
can convert to each other at certain physical conditions.
At high temperature, HNC can be reprocessed into HCN through hydrogen 
exchange reaction. Hence, the higher HNC abundance in \object{CRL\,3068}
might be related to its extreme cold environment. Furthermore,
HNC can be destructed by atomic oxygen through the reaction

\begin{equation}
{\rm HNC+O \rightarrow NH+CO}.
\end{equation}

Therefore, the higher HNC/HCN abundance ratio probably reflects the extremely carbon-rich nature of \object{CRL\,3068}.  \citet{schilke92} suggest that the HNC/HCN abundance increases with increasing dust depletion factor because reaction~(1) can be suppressed as a result of freeze-out of heavy elements onto dust grains.  If this is the case,  the extreme carbon star \object{CRL\,3068} may have a larger dust depletion factor than  \object{IRC+10216}.  
Our observations also suggest that the column density ratio of HNC/HCN in  \object{CRL\,3068} is about one order magnitude  lower than that in PNs \citep{josselin03}. This is consistent with the observations of \citet{herpin02} who found that HNC is enhanced with respect to HCN at post-AGB stage due to certain
chemical processes in the photodissociation region.

The chemical processes of CN are interconnected with those of HCN and HNC.  CN can be produced through photodissociation of HCN or HNC. On the other hand, CN can be reprocessed into HCN or HNC through reaction with molecular hydrogen. Very efficient  photodissociation has been found in the highly evolved AGB envelope \object{CIT\,6} \citep{zhang09}.  In  \object{CRL\,3068}, we do not find evidence showing enhancement of CN. 
C$_2$H and HC$_3$N can be reprocessed by the same chemical precursor, C$_2$H$_2^+$. The abundance ratio of the two species can provide a test for the chemical formation path \citep{wootten80}. We found that the abundance ratio $f$(C$_2$H)/$f$(HC$_3$N) in \object{CRL\,3068} is 7.7, lying within the range of 3--10 found by \citet{wootten80} in a wide variety of interstellar clouds and being consistent with gas-phase reaction scheme.

Among the most interesting finding in this study is the discovery of the cyclic molecule $c$-C$_3$H in  \object{CRL\,3068}.  This follows the detection of another cyclic molecule C$_3$H$_2$ in this source by \citet{wannier91}.  We further note that HC$_3$N and CH$_3$CN seem to be slightly enhanced in \object{CRL\,3068}. In the proto-planetary nebula \object{CRL\,618}, cyanopolyynes chains are formed through the polymerization of HCN \citep{pardo05}.  Our observations lead us to believe that  \object{CRL\,3068} may be more evolved than \object{IRC+10216} and  cyclic and longer chain molecules are quickly being synthesized in its envelope.
Since aromatic molecules emerge shortly after the end of the AGB \citep{kwo04}, the possible role that ring and chain molecules play in the synthesis of aromatic compounds is an important question in the study of circumstellar chemistry.

The interaction between stellar winds from AGB stars usually forms shock waves.
\citet{woods03} have found that shocks alter the chemical compositions (e.g. CS and SiO) in the inner regions of some AGB envelopes. According to a non-equilibrium chemical model presented by \citet{willacy98}, shocks can also strongly enhance SiO, which  is a commonly observed species in carbon rich envelopes.
However, we do not find enhancement of SiO in  \object{CRL\,3068}, probably suggesting 
that freeze out onto dust grains is significant in the extremely carbon rich envelope.

\section{Summary and conclusions} 

In this  molecular line survey toward the extreme carbon star \object{CRL~3068} in the $\lambda$ 1.3\,mm and $\lambda$ 2\,mm windows, 72 molecular lines from 23 species were detected with three lines remain unidentified. The spectra are dominated by carbon-bearing molecules. The species $c$-C$_3$H, CH$_3$CN, SiC$_2$, C$^{17}$O and C$^{18}$O, HC$^{15}$N, HN$^{13}$C, C$^{33}$S, C$^{34}$S, $^{13}$CS, $^{29}$SiS, and 
$^{30}$SiS are detected for the first time in this object. 
From these observations, we have derived the chemical abundances and isotopic ratios of the molecular species. A comparison between these  observations with the spectra of other AGB envelopes obtained
previously using the same observations settings allows us to rigorously investigate the chemical processes in circumstellar envelopes.  
Compared to the archetypal carbon star \object{IRC+10216}, \object{CRL~3068} is more carbon rich and
shows an enhancement of $c$-C$_3$H and HN$^{13}$C.
The overall assessment is that \object{CRL~3068} is a more evolved object on the AGB than IRC+10216. However, it should bear in mind that our conclusions
are mainly based upon the comparison of relative line intensities which
are assumed to directly correspond to the molecular abundances. To obtain more
reliable absolute abundances and to make stronger statements about the 
chemical evolution in circumstellar envelops, interferometric observations and 
sophisticated treatments of radiative transfer processes are required.

This study is a part of a long-term project of investigating the evolution of circumstellar chemistry in the late stages of stellar evolution using the $\lambda$ 2\,mm and $\lambda$ 1.3\,mm spectra of a sample of AGB stars, PPNs, and PNs.  Such a systematic study of the gas-phase molecules presents an important step to understand the synthesis of more complex aromatic and aliphatic compounds in evolved stars, and their role in the distribution of organic compounds throughout the Galaxy \citep{kwo09}.  A detailed study of chemical processes in different physical conditions and evolutionary stages
will be reported in the future.

\acknowledgments

The 12\,m telescope and the SMT are operated by the Arizona Radio Observatory 
(ARO), Steward Observatory, University of Arizona. YZ wish to thank the ARO 
staff for their hospitality during his stay at Kitt Peak and Mt. Graham. JN acknowledges financial support from Seed Funding Program for Basic Research in HKU (200802159006).  The work was supported by the Research Grants Council of the Hong Kong under grants HKU7020/08P and HKU7033/08P.
We also thank the anonymous referee for many helpful comments.

\appendix

\section{Olofsson formula for molecular abundance}

For the reader's convenience, we deduce
the formula presented by \citet{olofsson96} for the calculation
of molecular abundances respect to H$_2$. The basic assumptions
include: 1) all molecular emission originates from a spherical shell with inner 
radius $r_i$ and outer radius $r_e$; 2) the expansion velocity $v_e$,
the mass loss rate $\dot{M}$, and the excitation temperature
$T_{\rm ex}$ are constant along the radius 3) the molecular abundance respect 
to H$_2$ is homogeneous in the envelope.

For a Gaussian beam measuring a molecular line, the main beam temperature 
at a given velocity is obtained by
\begin{equation}
T_R(v)=\frac{\int \exp{(-\frac{4\ln2}{\theta_b^2}p^2)}T_s(v,p)d\Omega}{\int \exp{(-\frac{4\ln2}{\theta_b^2}p^2)}d\Omega}\\
=\frac{8\ln2}{\theta_b^2}\int \exp{(-\frac{4\ln2}{\theta_b^2}p^2)}T_s(v,p)pdp,
\end{equation}
where $p$ is the angular distance to the center, $\theta_b$ is the
half-power beam width, and $T_s(v,p)$ is the source brightness temperature at 
$v$ and $p$. For a given volume element at a distance to the center of the
envelope $r$, we have
\begin{equation}
p=\frac{r}{D}\sqrt{1-\left({v}/{v_e}\right)^2},
\end{equation}
where  D is the distance between the antenna and
 the source and $v$ is the velocity along the line of sight.
From the radiative transfer equation, the source brightnesss
temperature is given by
\begin{equation}
T_s(v,p)=T_{\rm ex}[1-e^{-\tau(v,p)}],
\end{equation}
where $\tau(v,p)$ is the optical depth. Substituting Equ.~A2 and Equ.~A3
into Equ.~A1, we have 
\begin{equation}
T_R(v)=8\ln2 T_{\rm ex}
\int^{x_e}_{x_i} e^{-{4\ln2}[1-(v/v_e)^2]x^2}
\left[1-\left({v}/{v_e}\right)^2\right][1-e^{-\tau(v,x)}]xdx,
\end{equation}
where we have defined   $x={r}/({\theta_bD})$
and thus $x_{i,e}={r_{i,e}}/({\theta_bD})$.

Accroding to radiative transfer equation,
$\tau(v,x)$ is given by
\begin{equation}
\tau(v,x)=
\frac{c^3g_uA_{ul}n(x)}{8\pi  \delta v\nu^3Q(T_{\rm ex})}
(1-e^{-h\nu/kT_{\rm ex}})e^{-E_l/kT_{\rm ex}}
\delta z,
\end{equation}
where $g_u$ is the weight of the upper level, $A_{ul}$ the Einstein
coefficient, $h$ the Planck constant, $k$ the
Boltzman constant, $\nu$ the line frequency, $n(x)$ the number
density of the species, $Q(T_{\rm ex})$ the partition function,
$E_l$ the energy of lower level,
and $\delta z$ the path length element along the line of sight.
From the relation $z=r(v/v_e)=Dp[1-(v/v_e)^2]^{-0.5}(v/v_e)$,
we have
\begin{equation}
\delta z=\frac{Dp}{v_e}[1-(v/v_e)^2]^{-1.5}\delta v
 =\frac{r}{v_e}[1-(v/v_e)^2]^{-1}\delta v.
\end{equation}
From Equs,~A5 and A6, we obtain
\begin{equation}
{\tau(v,x)}=
\frac{n(x)x}{n(1)}
\left[1-\left({v}/{v_e}\right)^2\right]^{-1}\tau(0,1),
\end{equation}
where
\begin{equation}
{\tau(0,1)}\approx
\frac{hc^3g_uA_{ul}n(1)\theta_bD}{8\pi k\nu^2v_eT_{\rm ex}Q(T_{\rm ex})}
e^{-E_l/kT_{\rm ex}}.
\end{equation}
The number density for a molecular species is given by
$$
n(r)=\frac{f_{\rm X}\dot{M}_{\rm H_2}}{m_{\rm H_2}v_er^2},
$$
or
\begin{equation}
n(x)=\frac{f_{\rm X}\dot{M}_{\rm H_2}}{m_{\rm H_2}v_e}\frac{1}{(x\theta_bD)^2}.
\end{equation}
where $f_{\rm X}$ is the abundance relative to H$_2$, $m_{\rm H_2}$
is the mass of a hydrogen molecule, and $\dot{M}_{\rm H_2}$ 
is the mass loss rate for H$_2$. Substituting Equs.~A7--A9 into A4 and
assuming $\tau(v,x)<<1$, we obtain
the molecular abundance relative to H$_2$
\begin{equation}
f_{\rm X}\approx
\frac{\pi}{2\ln2}
\frac{ km_{\rm H_2}}{hc^3}
\frac{v_e\theta_bD}{\dot{M}_{\rm H_2}}
\frac{Q(T_{\rm ex})\nu_{ul}^2}{g_uA_{ul}}
\frac{e^{E_l/kT_{\rm ex}}\int T_R(v)dv}{\int^{x_e}_{x_i}e^{-4x^2\ln2}dx}.
\end{equation}

\begin{figure*}
\epsfig{file=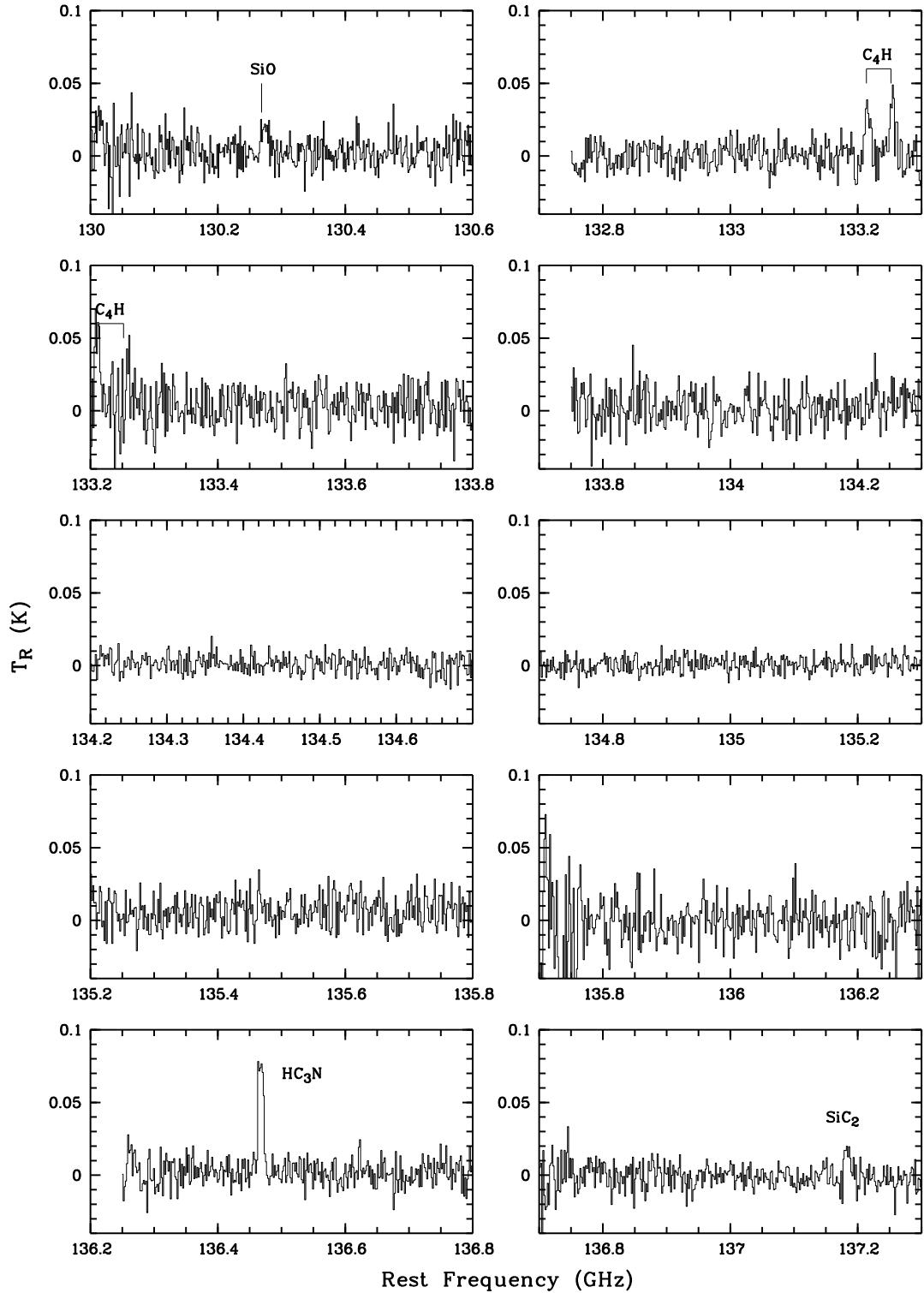,
height=20cm, }
\caption{Spectrum of CRL\,3068 in the frequency range
130--163\,GHz obtained with the ARO 12\,m telescope.
The spectra have been smoothed to a frequency resolution of 1\,MHz.
}
\label{spe_12m}
\end{figure*}

\addtocounter{figure}{-1}
\begin{figure*}
\epsfig{file=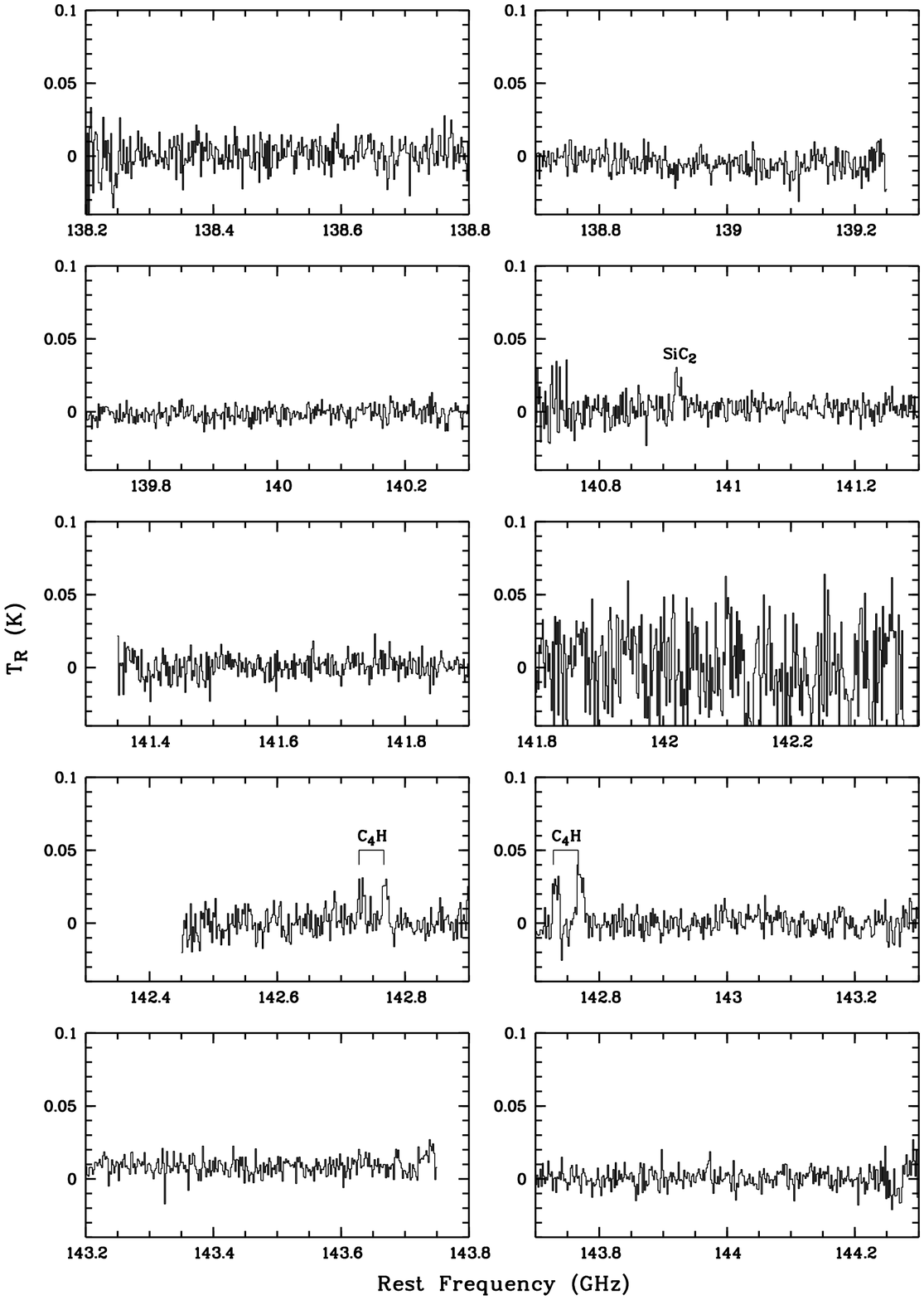,
height=20cm, }
\caption{continued.}
\end{figure*}

\addtocounter{figure}{-1}
\begin{figure*}
\epsfig{file=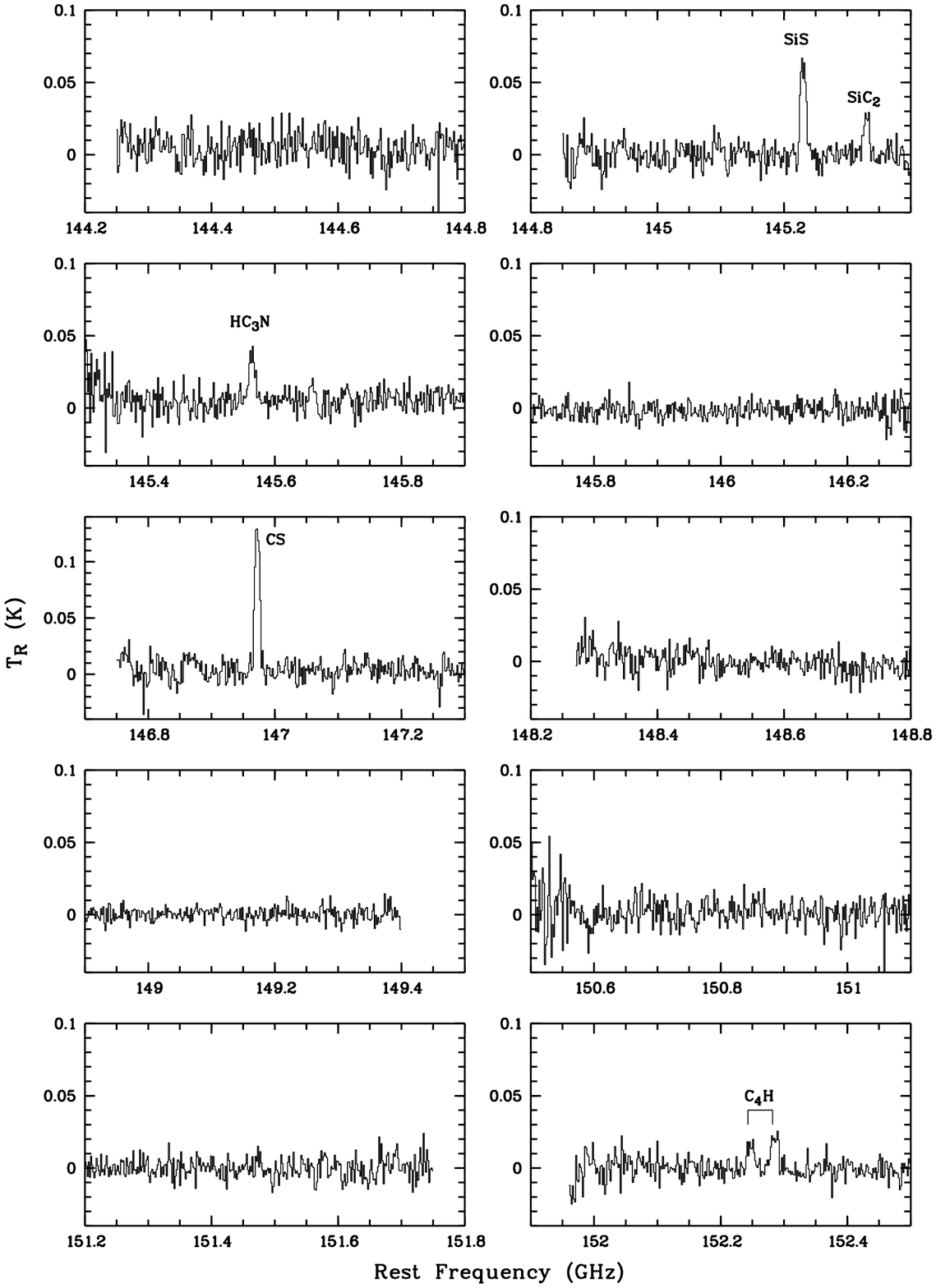,
height=20cm, }
\caption{continued.}
\end{figure*}

\addtocounter{figure}{-1}
\begin{figure*}
\epsfig{file=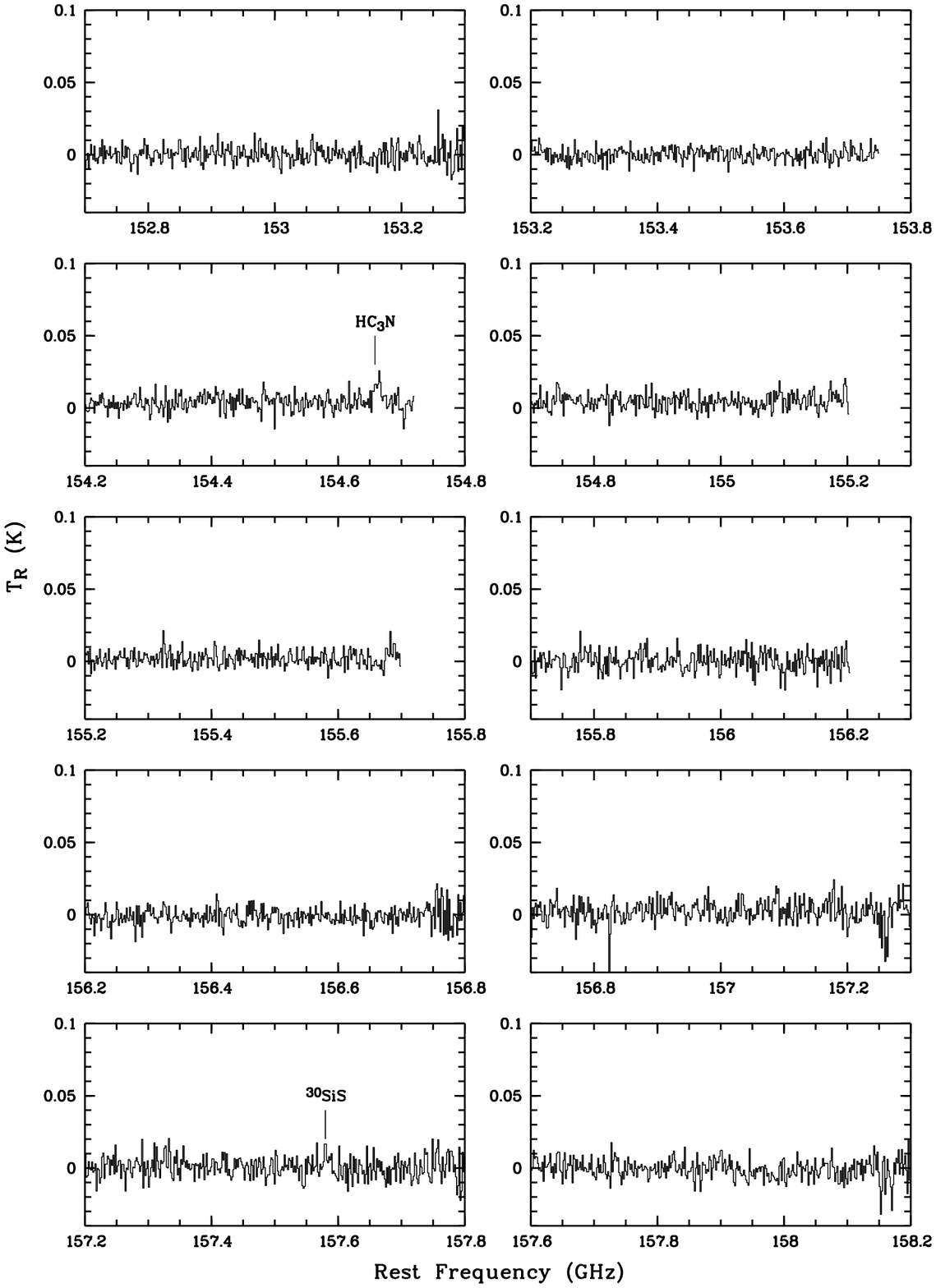,
height=20cm, }
\caption{continued.}
\end{figure*}

\addtocounter{figure}{-1}
\begin{figure*}
\epsfig{file=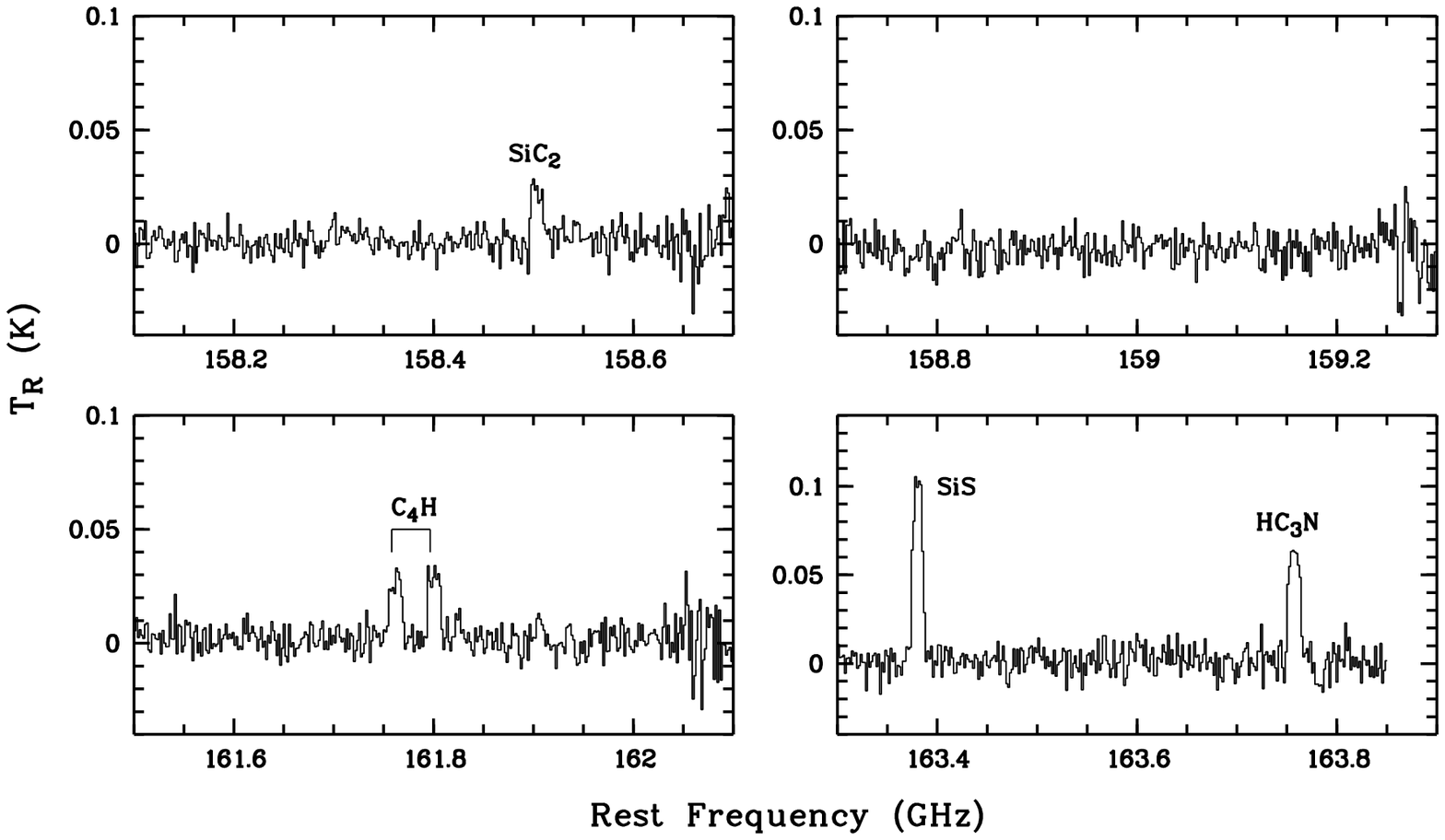,
height=8cm, }
\caption{continued.}
\end{figure*}

\begin{figure*}
\epsfig{file=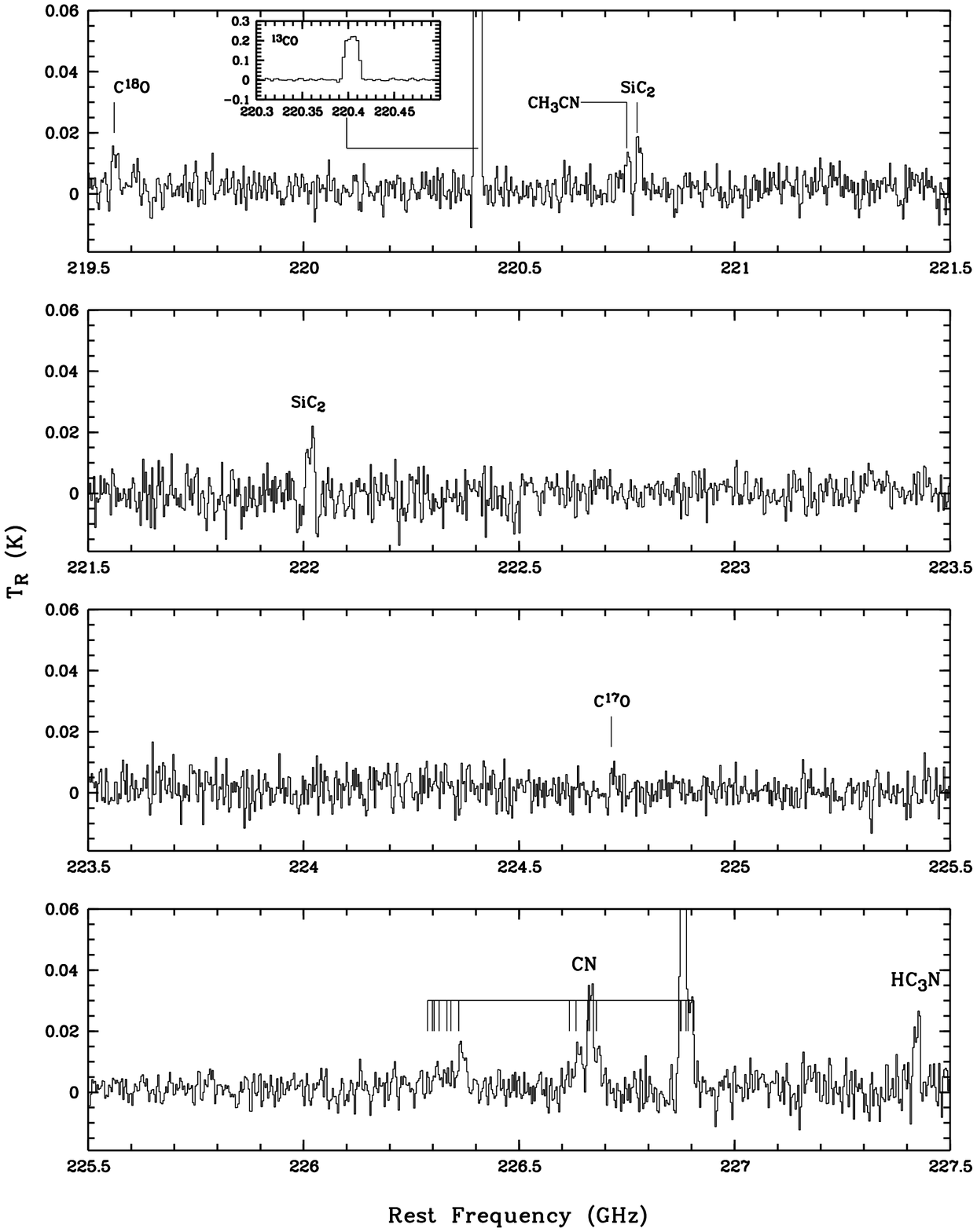,
height=20cm, }
\caption{Spectrum of CRL\,3068 in the frequency ranges
219.5--267.5\,GHz
obtained with the SMT 10\,m telescope.
The spectra have been smoothed to a frequency resolution of 3\,MHz.
}
\label{spe_smt}
\end{figure*}

\addtocounter{figure}{-1}
\begin{figure*}
\epsfig{file=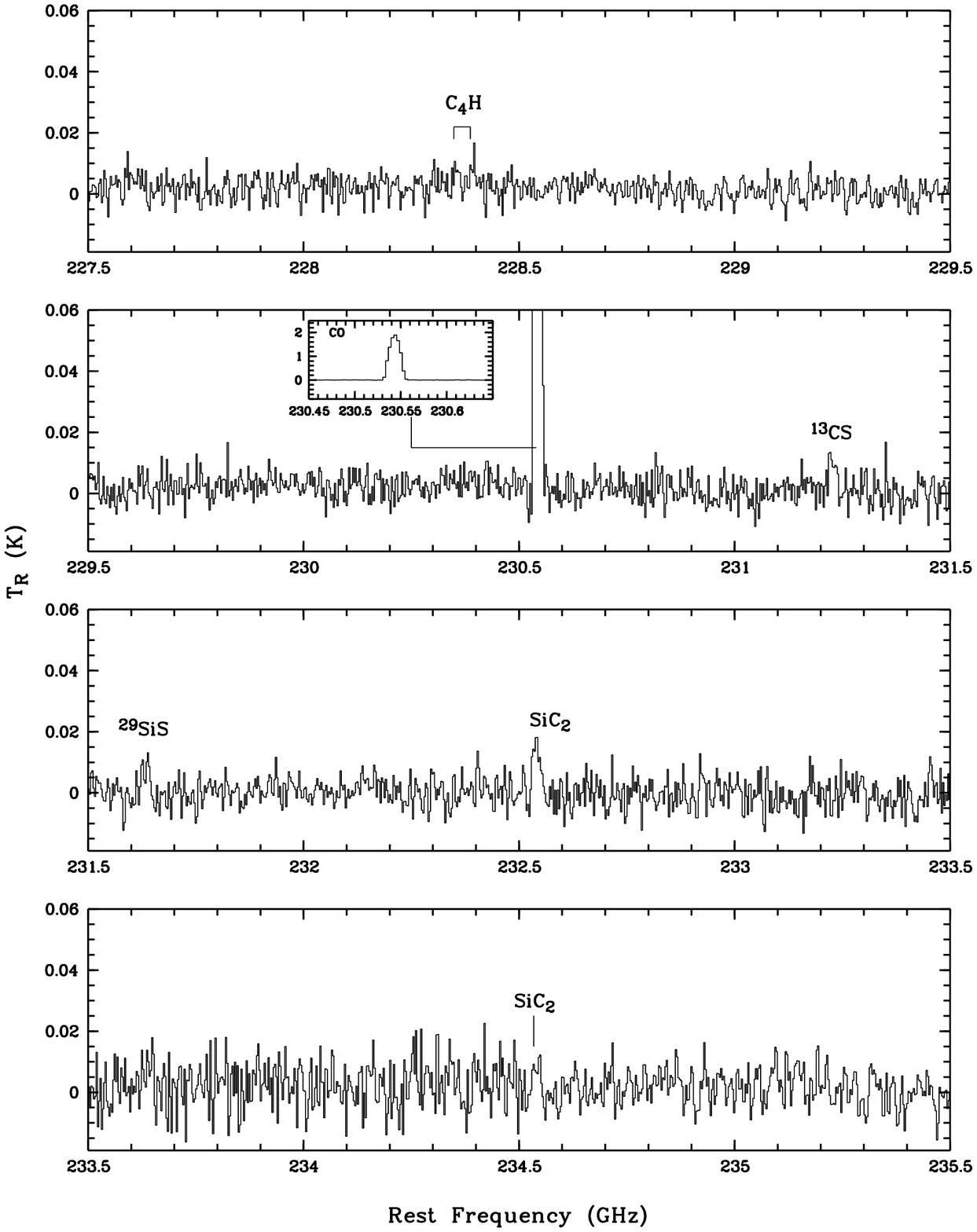,
height=20cm, }
\caption{continued.}
\end{figure*}

\addtocounter{figure}{-1}
\begin{figure*}
\epsfig{file=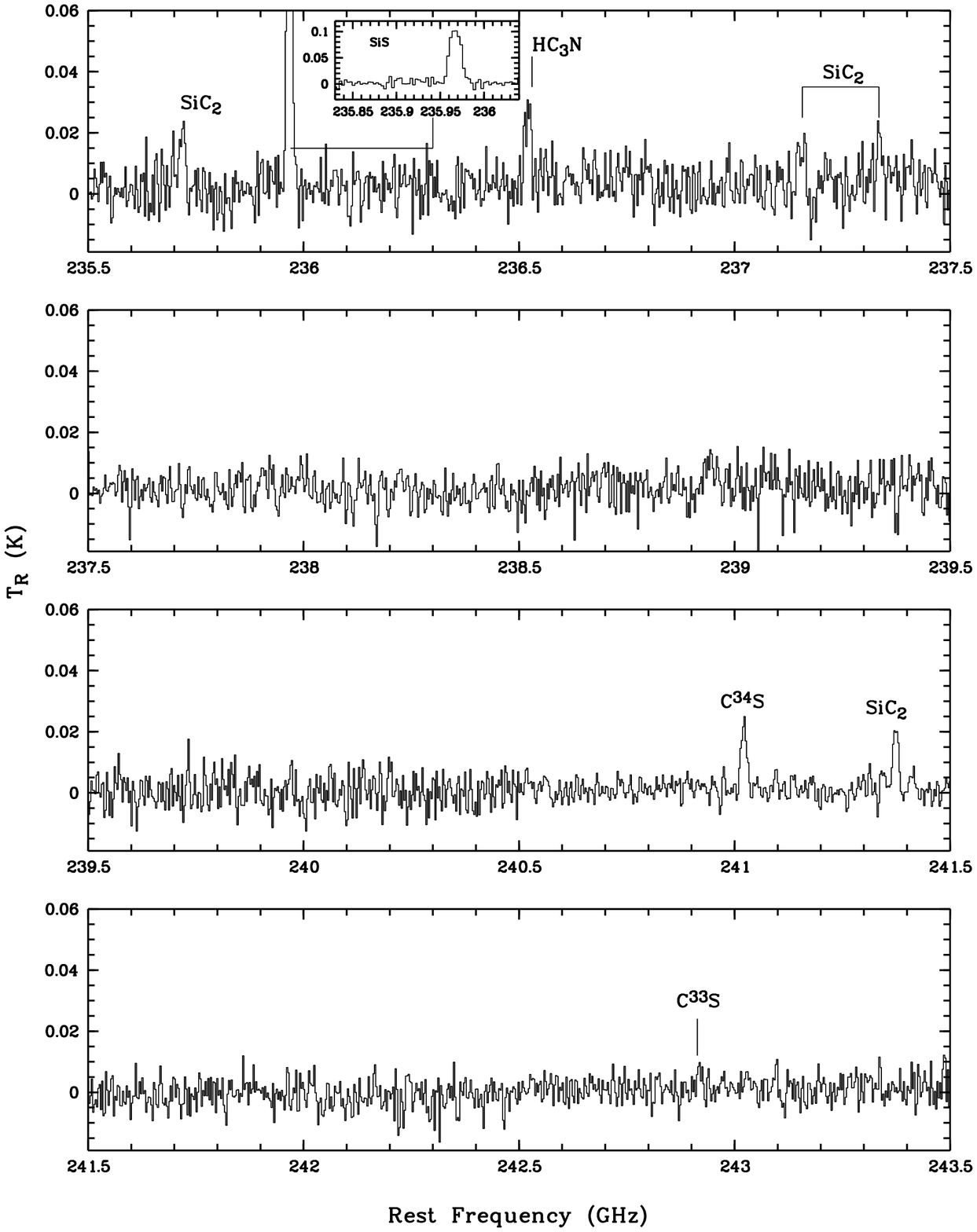,
height=20cm, }
\caption{continued.}
\end{figure*}

\addtocounter{figure}{-1}
\begin{figure*}
\epsfig{file=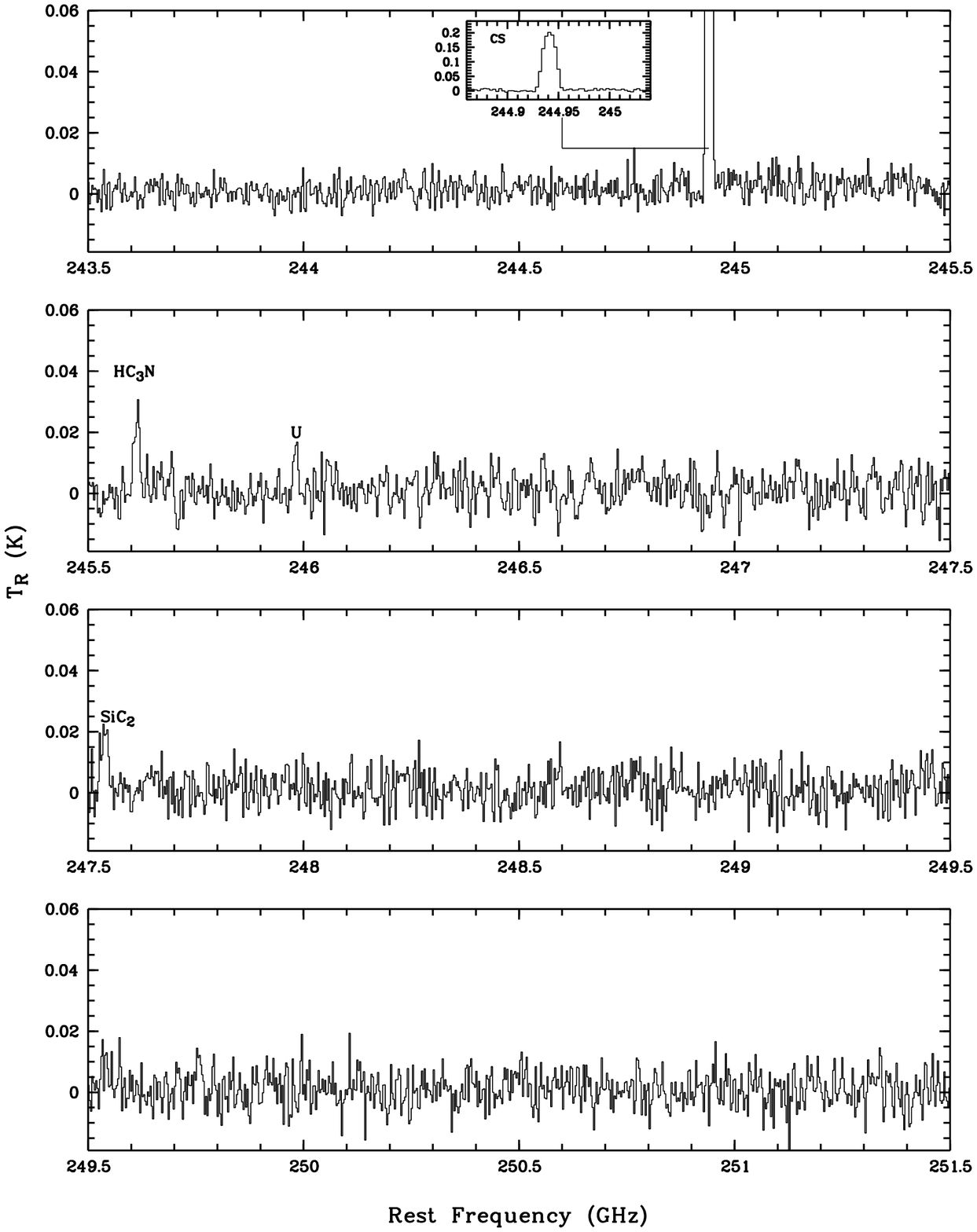,
height=20cm, }
\caption{continued.}
\end{figure*}

\addtocounter{figure}{-1}
\begin{figure*}
\epsfig{file=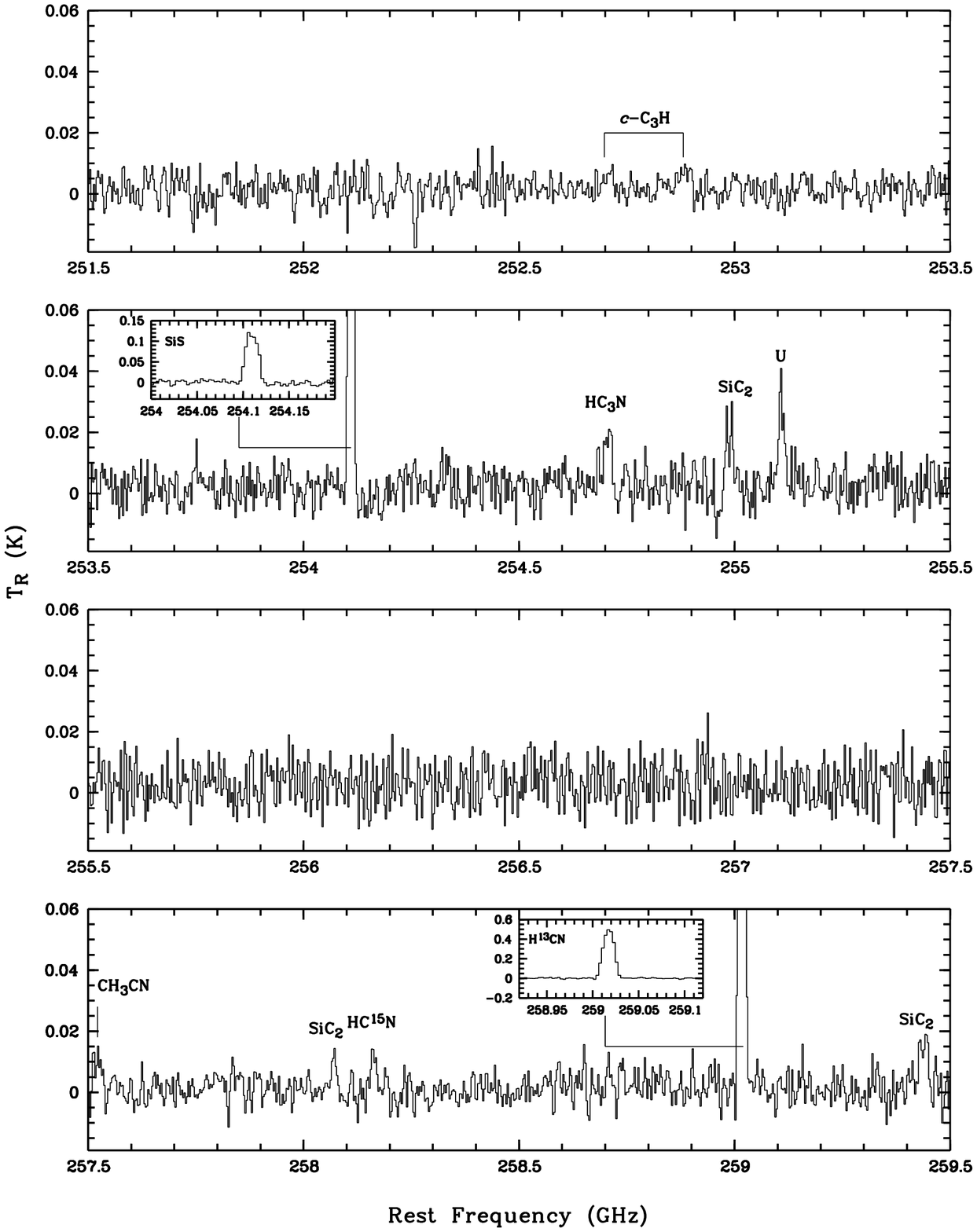,
height=20cm, }
\caption{continued.}
\end{figure*}

\addtocounter{figure}{-1}
\begin{figure*}
\epsfig{file=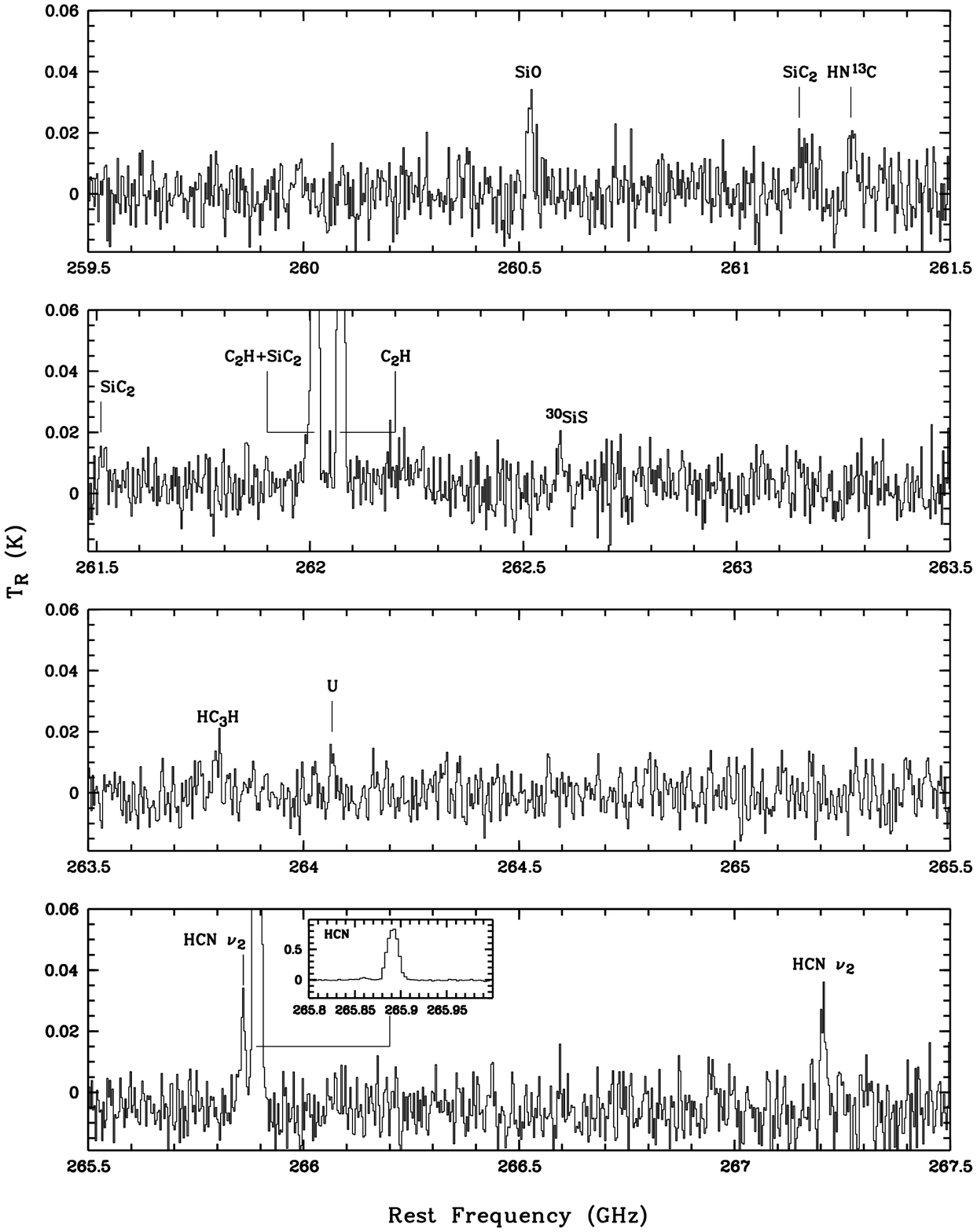,
height=20cm, }
\caption{continued.}
\end{figure*}

\begin{figure*}
\epsfig{file=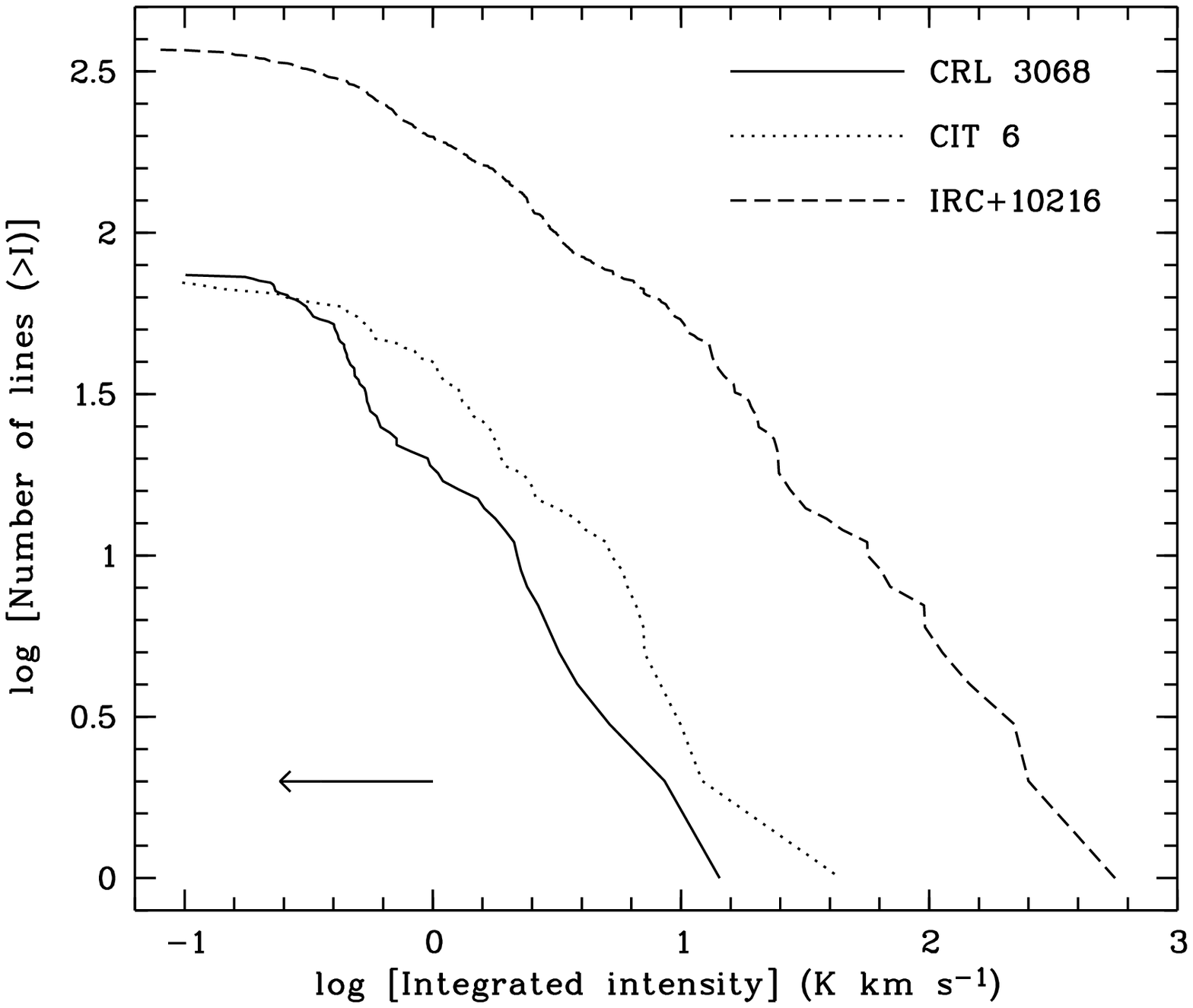,
height=12cm}
\caption{Cumulative number of observed lines exceeding 
a given value. 
The spectra of IRC+10216 and CIT~6 are 
from \citet{he08} and \citet{zhang09}, respectively.
All the spectra were obtained using the same telescope settings
in the same frequency ranges.
If the distance to an object is increased by a factor of two, the
curve would move toward left by a length denoted by the shaft of the 
arrow in the lower left-hand corner.
}
\label{cum}
\end{figure*}

\begin{figure*}
\epsfig{file=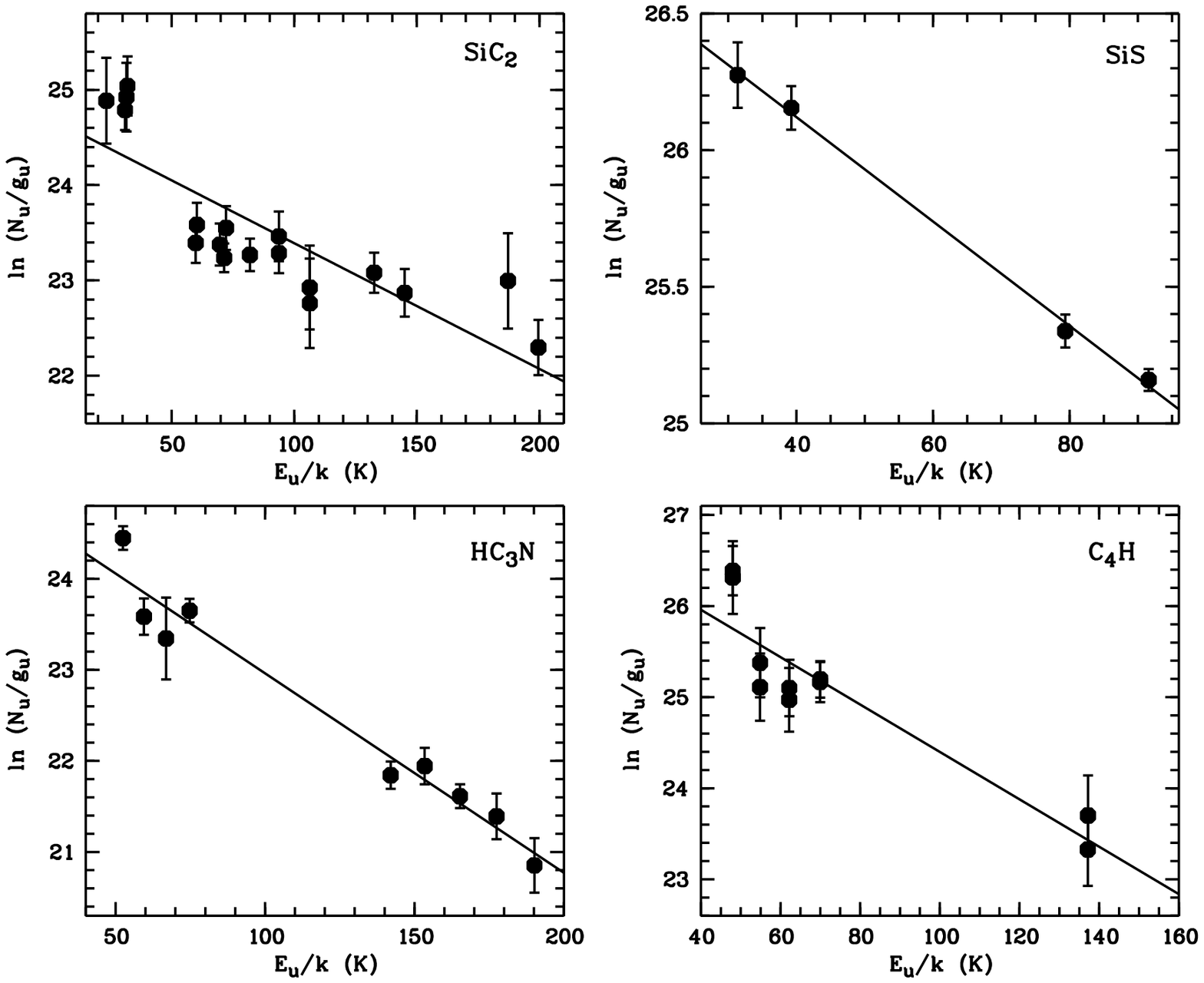,
height=12cm}
\caption{Rotational diagrams for the detected species in CRL\,3068.
}
\label{dia}
\end{figure*}

\begin{figure*}
\epsfig{ file=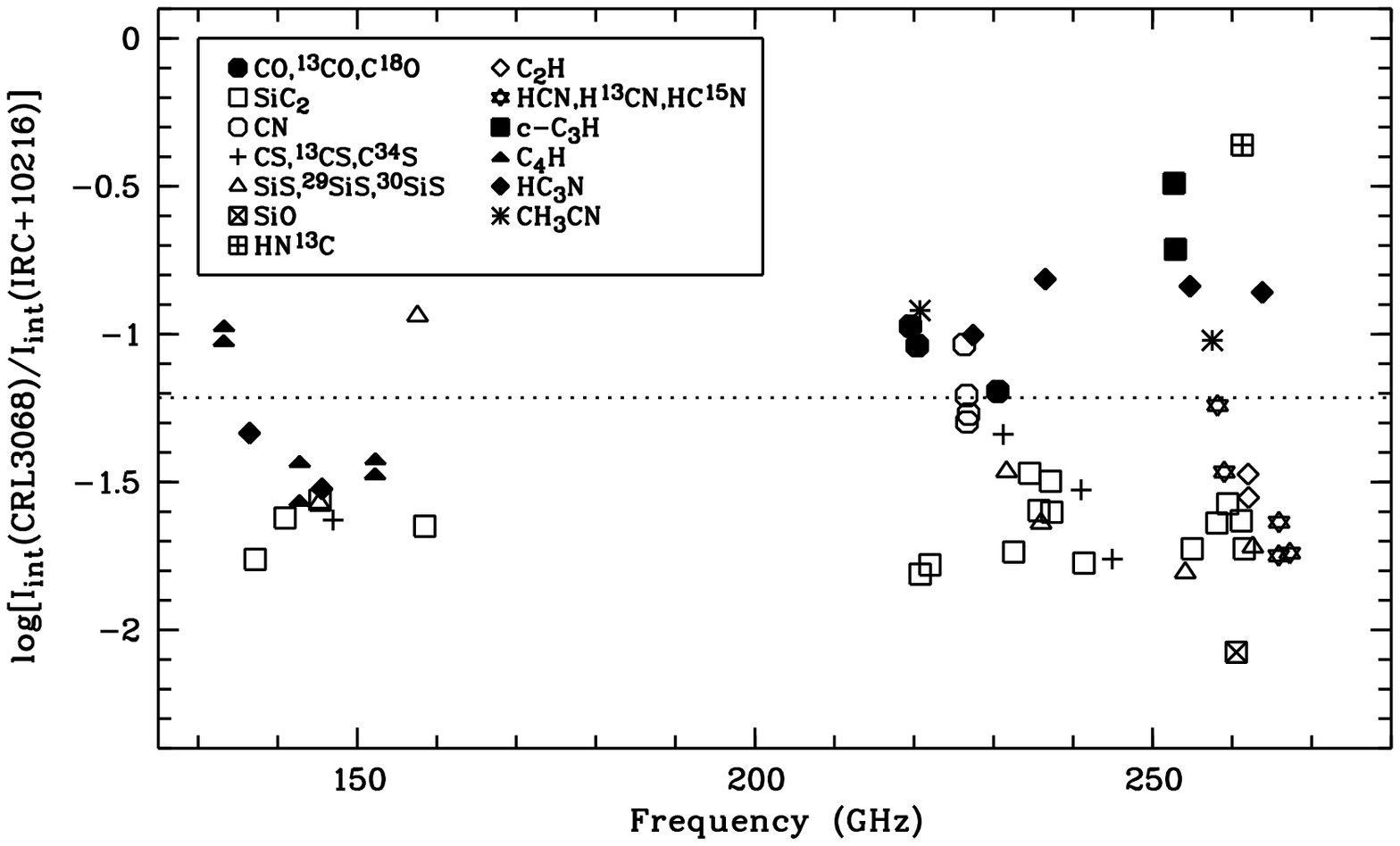,
height=10cm}
\caption{ Integrated intensity ratios of the lines detected
in CRL\,3068 and those detected in IRC+10216. The dotted line
represents the average value.
}
\label{linecomp}
\end{figure*}

\clearpage

\begin{deluxetable}{lllrcrc}
\tablecaption{Molecular transitions detected in  CRL\,3068.
\label{line}}
\tabletypesize{\scriptsize}
\tablewidth{0pt}
\tablehead{
\colhead{Species}& \colhead{Transition} & \colhead{Frequency$^a$} &  \colhead{$rms$}
&\colhead{$T_{\rm R}$} &\colhead{$\int T_{\rm R}$d$v$$^b$} &  \colhead{$\Delta V_{FWHM}$} \\
\colhead{(MHz)} & &\colhead{(upper--lower)} &\colhead{(mK)}  &\colhead{(K)} & \colhead{(K~km/s)}& \colhead{(km/s)}  \\
}
\startdata
 CO         & J=2--1               &230538.0&  4& 1.893&35.942& 17.7 \\
 $^{13}$CO  & J=2--1               &220398.7&  4& 0.221& 5.149& 21.2 \\
 C$^{17}$O  & J=2--1               &224714.4&  4& 0.009& 0.23:& ... \\
 C$^{18}$O  & J=2--1               &219560.4&  4& 0.016& 0.313& 25.0 \\
 CS         & J=3--2               &146969.0& 11& 0.143& 2.658& 18.4 \\
            & J=5--4               &244935.6&  4& 0.201& 3.821& 17.3 \\
 C$^{33}$S  & J=5--4               &242913.6&  3& 0.010& 0.17:& ... \\
 C$^{34}$S  & J=5--4               &241016.1&  3& 0.025& 0.485& 21.2 \\
 $^{13}$CS  & J=5--4               &231221.0&  4& 0.013& 0.325& 33.5 \\
 C$_2$H     & N=3--2 J=7/2--5/2    &262005.3\$ &  7& 0.130& 3.233& 22.3\\
            & N=3--2 J=5/2--3/2    &262066.1&  7& 0.079& 1.950& 31.6 \\
 C$_4$H     & N=14--13 a           &133213.7& 14& 0.052& 1.274&  ... \\ 
            & N=14--13 b           &133252.1& 14& 0.035& 1.100&  ... \\
            & N=15--14 a           &142728.8& 11& 0.030& 0.452& 27.2 \\ 
            & N=15--14 b           &142767.3& 11& 0.029& 0.553& 27.3 \\
            & N=16--15 a           &152243.6&  9& 0.026& 0.459& 23.3 \\
            & N=16--15 b           &152282.1&  9& 0.029& 0.508& 27.7 \\
            & N=17--16 a           &161758.1&  8& 0.036& 0.715& 20.8 \\
            & N=17--16 b           &161796.6&  8& 0.041& 0.716& 22.2 \\
            & N=24--23 a           &228348.6&  4& 0.010& 0.26:& ...  \\
            & N=24--23 b           &228387.0&  4& 0.009& 0.32:& ...  \\
 $c$-C$_3$H &5(1,4)--4(1,3) 11/2--9/2  &252697.3&4& 0.010& 0.23:& ... \\
            &5(1,4)--4(1,3) 9/2--8/2   &252881.0&4& 0.010& 0.31:& ... \\
 CH$_3$CN   &J$_{\rm K}$=$12_1$--11$_1$&220743.0*&4& 0.014& 0.291& 13.2\\
            &J$_{\rm K}$=$12_0$--11$_0$&220747.2*&4&   ---&   ---&  ---\\
            &J$_{\rm K}$=$14_2$--$13_2$&257507.6*&4& 0.015& 0.264& 26.0\\
            &J$_{\rm K}$=$14_1$--$13_1$&257522.4*&4&   ---&   ---&  ---\\
            &J$_{\rm K}$=$14_0$--$13_0$&257527.4*&4&   ---&   ---&  ---\\
 CN         &N(J,F)=2(3/2,1/2)--1(3/2,1/2)&226287.4*& 4&0.010&0.20:&   ...\\
            &N(J,F)=2(3/2,1/2)--1(3/2,3/2)&226298.9*& 4&  ---&  ---&   ---\\
            &N(J,F)=2(3/2,3/2)--1(3/2,1/2)&226303.0*& 4&  ---&  ---&   ---\\
            &N(J,F)=2(3/2,3/2)--1(3/2,3/2)&226314.5*& 4&  ---&  ---&   ---\\
            &N(J,F)=2(3/2,3/2)--1(3/2,5/2)&226332.5*& 4&0.015&0.594&  22.5\\
            &N(J,F)=2(3/2,5/2)--1(3/2,3/2)&226341.9*& 4&  ---&  ---&   ---\\
            &N(J,F)=2(3/2,5/2)--1(3/2,5/2)&226359.9*& 4&  ---&  ---&   ---\\
            &N(J,F)=2(3/2,1/2)--1(1/2,3/2)&226616.6*& 4&0.016&0.405&  23.8\\
            &N(J,F)=2(3/2,3/2)--1(1/2,3/2)&226632.2*& 4&  ---&  ---&   ---\\
            &N(J,F)=2(3/2,5/2)--1(1/2,3/2)&226659.6*& 4&0.036&1.049&  21.8\\
            &N(J,F)=2(3/2,1/2)--1(1/2,1/2)&226663.7*& 4&  ---&  ---&   ---\\
            &N(J,F)=2(3/2,3/2)--1(1/2,1/2)&226679.3*& 4&  ---&  ---&   ---\\
            &N(J,F)=2(5/2,5/2)--1(3/2,3/2)&226874.2*& 4&0.088&2.399&  21.9\\
            &N(J,F)=2(5/2,7/2)--1(3/2,5/2)&226874.8*& 4& ---&  ---&   ---\\
            &N(J,F)=2(5/2,3/2)--1(3/2,1/2)&226875.9*& 4& ---&  ---&   ---\\
            &N(J,F)=2(5/2,3/2)--1(3/2,3/2)&226887.4*& 4& ---&  ---&   ---\\
            &N(J,F)=2(5/2,5/2)--1(3/2,5/2)&226892.1*& 4& ---&  ---&   ---\\
            &N(J,F)=2(5/2,3/2)--1(3/2,5/2)&226905.4*& 4& ---&  ---&   ---\\
 HCN        & J=3--2                 &265886.4&  7& 0.831&14.315& 15.8 \\
            & $\nu_2$=1$^{1e}$ J=3--2&265852.8&  7& 0.034& 0.401& 11.5 \\
            & $\nu_2$=2$^0$ J=3--2   &267243.2&  7& 0.036& 0.231& 14.5 \\
 HC$^{15}$N & J=3--2                 &258154.7&  4& 0.014& 0.233& 22.3 \\
 H$^{13}$CN & J=3--2                 &259011.8&  5& 0.495& 8.582& 15.8 \\
 HN$^{13}$C & J=3--2                 &261263.4&  7& 0.021& 0.445& 23.7 \\
 HC$_3$N    & J=15--14               &136464.4& 11& 0.082& 1.787& 21.4 \\
            & J=16--15               &145560.9& 10& 0.053& 0.953& 21.4 \\
            & J=17--16               &154657.3&  9& 0.020& 0.45:& ...  \\
            & J=18--17               &163753.4& 10& 0.075& 1.611& 21.8 \\
            & J=25--24               &227418.9&  4& 0.027& 0.530& 20.5 \\
            & J=26--25               &236512.8&  6& 0.031& 0.672& 25.3 \\
            & J=27--26               &245606.3&  4& 0.031& 0.541& 21.3 \\
            & J=28--27               &254699.5&  5& 0.021& 0.485& 19.3 \\
            & J=29--28               &263792.3&  6& 0.021& 0.310& 26:  \\
 SiC$_2$    & 6$_{0,6}$--5$_{0,5}$   &137180.8& 10& 0.022& 0.424& 21:  \\
            & 6$_{2,5}$--5$_{2,4}$   &140920.1&  9& 0.025& 0.483& 20:  \\
            & 6$_{2,4}$--5$_{2,3}$   &145325.8&  9& 0.029& 0.560& 20.6 \\
            & 7$_{0,7}$--6$_{0,6}$   &158499.2&  7& 0.033& 0.617& 25.0 \\
            &10$_{0,10}$--9$_{0,9}$  &220773.7&  4& 0.019& 0.380& 18.3 \\
            & 9$_{2,7}$--8$_{2,6}$   &222009.4&  5& 0.022& 0.400& 29.0 \\
            &10$_{2,9}$--9$_{2,8}$   &232534.1&  4& 0.018& 0.411& 29.7 \\
            &10$_{8,2}$--9$_{8,1}$   &234534.0*&  6& 0.012& 0.222& 29.3\\
            &10$_{8,3}$--9$_{8,2}$   &234534.0*&  6&   ---&   ---&  ---\\
            &10$_{6,5}$--9$_{6,4}$   &235713.0*&  5& 0.024& 0.415&  15:\\
            &10$_{6,4}$--9$_{6,3}$   &235713.1*&  5&   ---&   ---&  ---\\
            &10$_{4,7}$--9$_{4,6}$   &237150.0&  5& 0.019& 0.417&  30: \\
            &10$_{4,6}$--9$_{4,5}$   &237331.3&  5& 0.024& 0.35:&  15: \\
            &11$_{0,11}$--10$_{0,10}$&241367.7&  3& 0.020& 0.439& 18.6 \\
            &10$_{2,8}$--9$_{2,7}$   &247529.1&  5& 0.022& 0.544& 22.8 \\
            &11$_{2,10}$--10$_{2,9}$ &254981.5&  5& 0.030& 0.466&  28: \\
            &11$_{8,3}$--10$_{8,2}$  &258065.0*&  4& 0.014& 0.189& 16.2\\
            &11$_{8,4}$--10$_{8,3}$  &258065.0*&  4&   ---&   ---&  ---\\
            &11$_{6,6}$--10$_{6,5}$  &259433.3*&  5& 0.019& 0.502& 23.0\\
            &11$_{6,5}$--10$_{6,4}$  &259433.3*&  5&   ---&   ---&  ---\\
            &11$_{4,8}$--10$_{4,7}$  &261150.7&  7& 0.016& 0.33:&  ... \\
            &11$_{4,7}$--10$_{4,6}$  &261509.3&  7& 0.015& 0.28:& 22.4\\
            &12$_{0,12}$--11$_{0,11}$&261990.7\$ &  7& 0.130& 3.233& 22.3\\
 SiO        & J=3--2                 &130268.6& 12& 0.026& 0.976& 22.1 \\
            & J=6--5                 &260518.0&  7& 0.034& 0.537& 15.7 \\
 SiS        & J=8--7                 &145227.0&  9& 0.078& 1.517& 20.3 \\
            & J=9--8                 &163376.7& 10& 0.119& 2.127& 19.5 \\ 
            & J=13--12               &235961.1&  6& 0.101& 2.188& 18.8 \\
            & J=14--13               &254102.9&  5& 0.121& 2.265& 18.4 \\
 $^{29}$SiS & J=13--12               &231626.7&  4& 0.013& 0.243& 30:  \\ 
 $^{30}$SiS & J=9--8                 &157579.7&  9& 0.025& 0.44:& ...  \\
            & J=15--14               &262585.0&  7& 0.021& 0.101&19.4  \\
          U &                        &245982.0&  5& 0.017& 0.605&14.7  \\
            &                        &255108.0&  5& 0.041& 0.817&16.0  \\
            &                        &264067.0&  5& 0.016& 0.175&13.9  \\
\enddata
\tablenotetext{{\it a}}{ $*$--unsolved hyperfine structure lines; \$ blended
with other species.}
\tablenotetext{{\it b}}{ The colon indicates uncertain detections.
The pointing uncertainties induced errors are up to 20\%--30$\%$.}
\end{deluxetable}

\begin{deluxetable}{lccrccr}
\tablecaption{Excitation temperatures, column densities
and abundances respect to H$_2$$^a$.
\label{abundance}}
\tabletypesize{\scriptsize}
\tablewidth{0pt}
\tablehead{
\colhead{Species} & \colhead{$T_{\rm ex}$\,(K)$^b$} & \multicolumn{2}{c}{$N$\,(cm$^{-2}$)}&
& \multicolumn{2}{c}{$f_{\rm X}$}\\
\cline{3-4} \cline{6-7}
& &\colhead{This paper$^c$} & \colhead{F94$^d$} & &\colhead{This paper$^c$} & \colhead{W03$^d$}\\
}
\startdata
CO         & \nodata & 6.20(16) & \nodata         &&1.5(-4) &\nodata \\
$^{13}$CO  & \nodata & 1.17(16) & \nodata         &&2.7(-5) &\nodata \\
C$^{17}$O  & \nodata & 5.58(14):& \nodata         &&1.2(-6):&\nodata \\
C$^{18}$O  & \nodata & 7.66(14) & \nodata         &&1.7(-6) &\nodata \\
CS         & \nodata & 3.24(13) & \nodata         &&5.8(-8) &3.7(-7) \\
C$^{33}$S  & \nodata & 1.38(12):& \nodata         &&2.7(-9):&\nodata \\
C$^{34}$S  & \nodata & 3.95(12) & \nodata         &&7.8(-9) &\nodata \\
$^{13}$CS  & \nodata & 2.91(12) & \nodata         &&5.9(-9) &\nodata \\
C$_2$H     & \nodata & 1.82(14) & $1.5\pm1.1$(14) &&2.7(-7) &5.7(-6) \\
C$_4$H     & $38\pm18$      & 3.52(14) & 7.0(14)         &&6.5(-7) &\nodata \\
$c$-C$_3$H & \nodata & 1.78(13):& \nodata         &&3.6(-8):&\nodata \\
CH$_3$CN   & \nodata & 5.77(12) & \nodata         &&8.1(-8) & $<1.2$(-7)\\
CN         & \nodata & 1.93(14) & \nodata         &&3.9(-7) &4.6(-7)\\
HCN        & \nodata & 4.17(13) & 6.0(14)         &&4.5(-8) &6.3(-6)\\
HC$^{15}$N & \nodata & 5.49(11) & \nodata         &&1.0(-9) &\nodata \\
H$^{13}$CN & \nodata & 2.01(13) & \nodata         &&3.7(-8) &3.0(-8)\\
HN$^{13}$C & \nodata & 1.26(12) & \nodata         &&2.3(-9) &$<1.4$(-8)\\
HC$_3$N    & $46\pm15$     & 1.78(13) & $1.9\pm1.7$(14) &&3.5(-8)& 5.0(-7)\\
SiC$_2$    & $76\pm36$     & 3.49(13) & \nodata         &&3.8(-8)&$<7.5$(8)\\
SiO        & \nodata & 1.64(12) & \nodata         &&3.0(-9)& \nodata \\
SiS        & $52\pm16$      & 5.69(13) & $5.4\pm9.0$(13) &&1.5(-7)&6.7(-7)\\
$^{29}$SiS & \nodata & 6.38(12) & \nodata         &&1.3(-8)& \nodata\\
$^{30}$SiS & \nodata & 2.83(12) & \nodata         &&5.2(-9)& \nodata  \\
\enddata
\tablenotetext{{\it a}}{$x(y)$ represents $x\times10^y$;}
\tablenotetext{{\it b}}{A constant excitation temperature of 53\,K was
assumed for the species for which
the rotation diagrams cannot be obtained; }
\tablenotetext{{\it c}}{For the species  with optically thick
emission (e.g. CO and HCN), this gives the lower limits;}
\tablenotetext{{\it d}}{F94: from Fukasaku et al. (1994); W03: form Woods et al. (2003).}
\end{deluxetable}

\begin{deluxetable}{llcccc}
\tablecaption{Isotopic abundance ratios.
\label{isoto}}
\tablewidth{0pt}
\tablehead{
\colhead{Isotopic ratio} & \multicolumn{2}{c}{CRL\,3068} & \colhead{CIT\,6$^a$} & \colhead{IRC+10216$^b$} & \colhead{Solar$^c$}\\
\cline{2-3}
& \colhead{Sepcies}& \colhead{Value}\\
}
\startdata
$^{12}$C/$^{13}$C &$^{12}$C$^{34}$S/$^{13}$C$^{32}$S & 29.7$^d$ &45.4  & 45       & 89   \\
                  &$^{12}$CO/$^{13}$CO               & 5.6$^e$&  \nodata  & \nodata       & \nodata \\
                  &$^{12}$CS/$^{13}$CS               & 9.8$^e$&  \nodata  & \nodata      & \nodata \\
                  &H$^{12}$CN/H$^{13}$CN             & 1.2$^e$&  \nodata  & \nodata       & \nodata \\
$^{14}$N/$^{15}$N  & H$^{13}$C$^{14}$N/H$^{12}$C$^{15}$N    &  1099$^f$& \nodata   &  \nodata       & 272 \\
                   & HC$^{14}$N/HC$^{15}$N            &  45$^e$ & \nodata   &  \nodata       & \nodata \\
$^{16}$O/$^{17}$O &$^{13}$C$^{16}$O/$^{12}$C$^{17}$O &668:$^f$& 890       &  967       & 2680 \\
                  &C$^{16}$O/C$^{17}$O               &125:$^e$& \nodata   &  \nodata       & \nodata \\
$^{16}$O/$^{18}$O &$^{13}$C$^{16}$O/$^{12}$C$^{18}$O & 472$^f$& \nodata   &  1172       & 499 \\
                  &C$^{16}$O/C$^{18}$O               & 88$^e$ & \nodata   &  \nodata       & \nodata \\
$^{17}$O/$^{18}$O &C$^{17}$O/C$^{18}$O               & 0.7:   &  \nodata  &  1.14       & 0.2   \\
$^{32}$S/$^{34}$S  &C$^{32}$S/C$^{34}$S              & 7.4$^e$&  6.7$^e$  & 18.9   & 22.5 \\
$^{33}$S/$^{34}$S  &C$^{33}$S/C$^{34}$S              & 0.3:   &  0.2:     & 0.19   & 0.18 \\
$^{29}$Si/$^{30}$Si& $^{29}$SiS/$^{30}$SiS           & 2.5    &   1.0    & 1.46    & 1.52 \\
$^{28}$Si/$^{30}$Si& $^{28}$SiS/$^{30}$SiS           & 28.8   &   8.8$^e$& 24.7    & 29.9 \\
$^{28}$Si/$^{29}$Si& $^{28}$SiS/$^{29}$SiS           & 11.5   &  8.9$^e$ & 17.2    & 19.6 \\
\enddata
\tablenotetext{{\it a}}{From Zhang et al. (2009);}
\tablenotetext{{\it b}}{From He et al. (2008) except the $^{12}$C/$^{13}$C ratio
which was taken from Cernicharo et al. (2000);}
\tablenotetext{{\it c}}{From Lodders (2003);}
\tablenotetext{{\it d}}{Assume that the $^{34}$S/$^{32}$S ratio is solar;}
\tablenotetext{{\it e}}{Should be treated as lower limits due to opacity effect;}
\tablenotetext{{\it f}}{Adopted: $^{12}$C/$^{13}$C=29.7.}
\end{deluxetable}

\end{document}